\documentclass[pra,superscriptaddress,twocolumn]{revtex4}
\usepackage{amsthm}
\usepackage{amsmath}
\usepackage{mathrsfs}
\usepackage{amssymb}
\usepackage{algorithm}
\usepackage{algpseudocode}
\usepackage{wasysym}
\usepackage{graphicx}
\usepackage{varwidth}
\usepackage{url}
\usepackage{dsfont}
\usepackage{float}
\usepackage{bm}
\usepackage{ifthen}
\usepackage[usenames,dvipsnames]{color}
\usepackage{mathrsfs}
\usepackage[colorlinks=true,citecolor=blue,urlcolor=black]{hyperref}
\usepackage{float}
\usepackage{afterpage}
\usepackage{subfigure}
\usepackage{microtype}
\usepackage{booktabs}

\usepackage[encapsulated]{CJK}
\newcommand{\sket}[1]{{\ensuremath{\lvert#1\rangle}}}
\newcommand{\lket}[1]{{\ensuremath{\left\lvert#1\right\rangle}}} 
\newcommand{\ket}[1]{\if@display\lket{#1}\else\sket{#1}\fi}
\newcommand{\sbra}[1]{{\ensuremath{\langle#1\rvert}}}
\newcommand{\lbra}[1]{{\ensuremath{\left\langle#1\right\rvert}}}
\newcommand{\bra}[1]{\if@display\lbra{#1}\else\sbra{#1}\fi}
\newcommand{\sbraket}[2]{{\ensuremath{\langle#1\rvert#2\rangle}}}
\newcommand{\lbraket}[2]{{\ensuremath{\left\langle#1\!\left\rvert\vphantom{#1}#2\right.\!\right\rangle}}}
\newcommand{\braket}[2]{\if@display\lbraket{#1}{#2}\else\sbraket{#1}{#2}\fi}

\newcommand{\sketbra}[2]{{\ensuremath{\lvert #1\rangle\!\langle #2\rvert}}}
\newcommand{\lketbra}[2]{{\ensuremath{\left\lvert #1\right\rangle\!\!\left\langle #2\right\rvert}}}
\newcommand{\ketbra}[2]{\if@display\lketbra{#1}{#2}\else\sketbra{#1}{#2}\fi}

\theoremstyle{plain}

\theoremstyle{definition}

\begin{document}
\begin{CJK*}{UTF8}{gbsn}
\title{Learnability and Complexity of Quantum Samples}
\author{Murphy Yuezhen Niu}
\affiliation{Google Research, 340 Main Street, Venice Beach, California 90291, USA}
\author{Andrew M. Dai}
\affiliation{Google Health,  Palo Alto, California 94304, USA}
\author{Li Li (李力)}
\affiliation{Google Research, 1600 Amphitheatre Parkway, Mountain View, California 94043, USA}
\author{Augustus Odena}
\affiliation{Google Research, 1600 Amphitheatre Parkway, Mountain View, California 94043, USA}
\author{Zhengli Zhao}
\affiliation{Google Research, 1600 Amphitheatre Parkway, Mountain View, California 94043, USA}
\author{Vadim Smelyanskyi}
\affiliation{Google Research, 340 Main Street, Venice Beach, California 90291, USA}
\author{Hartmut Neven}
\affiliation{Google Research, 340 Main Street, Venice Beach, California 90291, USA}
\author{Sergio Boixo}
\affiliation{Google Research, 340 Main Street, Venice Beach, California 90291, USA}

\begin{abstract}
 

Given a quantum circuit, a quantum computer can sample the output distribution exponentially faster in the number of bits than classical computers. A similar exponential separation has yet to be established in  generative models through quantum sample learning:  given samples from an $n$-qubit computation, can we learn the underlying quantum distribution using models with training parameters that scale polynomial in $n$ under a fixed training time? We study four kinds of generative models: Deep Boltzmann machine (DBM), Generative Adversarial Networks (GANs), Long Short-Term Memory (LSTM) and Autoregressive GAN, on learning quantum data set generated by deep random circuits. We demonstrate the leading performance of LSTM in learning quantum samples, and thus  the autoregressive structure present in the underlying quantum distribution from random quantum circuits.   Both numerical experiments and a theoretical proof in the case of the DBM show exponentially growing  complexity of learning-agent parameters required for achieving a fixed accuracy as $n$ increases. Finally, we establish a connection between learnability and the complexity of generative models by benchmarking learnability against different sets of samples drawn from probability distributions of variable degrees of complexities in their quantum and classical representations. 
 
\end{abstract}
  
\maketitle
\end{CJK*}

\section{Introduction}

	One of the main goals of quantum computing is to discover computational tasks where exponential quantum speed-ups exist. 
It has been conjectured~\cite{boixo2018characterizing,aaronson2016complexity,bouland2019complexity,movassagh2018efficient} and experimentally shown~\cite{supremacy2019quantum} that as the number of qubits $n$ in a sufficiently deep random quantum circuit increases, the amount of classical computation needed to sample from the corresponding distribution, given the quantum circuit, increases exponentially in $n$.  A similar exponential separation has not been established in a bottom-up generative task: given a set of samples produced by a quantum computation on $n$ qubits, can a classical agent learn the underlying distribution to a given accuracy using only polynomial-in-$n$ training parameters? In this work, we address this open question by studying the performance of various types of generative models, and study the relation between learnability and compression or complexity~\cite{david2016supervised,littlestone1986relating,ben2019learnability}. 


The state-of-the-art generative machine learning models are capable of modeling distributions of extremely high complexity. Applications of deep generative models range from image, audio, video and text synthesis to improving the physical modeling for Large Hadron Collider experiments~\cite{Nachman2018}. 
We apply four commonly used generative models, covering four main categories of generative methods, see Table~\ref{GenerativeArchitectureTable}. These include a Deep Boltzmann machine~(DBM) in form of exact construction with a known algorithm, see Appendix A; Long Short-Term Memory~(LSTM) through maximum likelihood optimization; Generative Adversarial Network~(GAN); and Sequential GAN~(SeqGAN) with recurrent network structure through min-max optimization over generative and discriminative networks. We show that autoregressivity is a critical feature to learn the distributions produced by the random circuits in \citet{supremacy2019quantum}: generative models with autoregressive structure such as LSTM~\citep{Hochreiter1997-mo} and SeqGAN~\cite{seqgan} outperform conventional GAN  in learning accuracy, and outperform both conventional GAN and DBM in the efficiency of training parameters for learning deep random quantum circuit outputs.  

\begin{table} 

\scalebox{0.9}{%

\begin{tabular}{ccc}
\toprule

 &  \multicolumn{1}{ c }{\bf ML} & \multicolumn{1}{ c }{ \bf GAN }  \\ 
\midrule
one-shot  & DBM   
      &   GAN \\
 autoregressive  &   LSTM
    &  SeqGAN  \\
\bottomrule
\end{tabular}} 

\caption{Generative models studied in this work with respect to their loss function (ML: maximum likelihood optimization; GAN: generative adversarial network cost minmax optimization) and generative process (one-shot: generate the whole sample in one shot; autoregressive: generate one bit after the other sequentially).}\label{GenerativeArchitectureTable}
\end{table}

DBMs are undirected graphical models realized through bipartite connections between adjacent layers of visible, hidden and deep layers. DBMs explicitly parameterize the dependencies in the joint   density model over the space of  inputs called visible layers. DBMs serve as powerful tools for    object or speech recognition~\cite{salakhutdinov2009deep,salakhutdinov2010efficient,srivastava2012multimodal}, as well as for representing quantum many-body wave functions~\cite{carleo2018constructing}. Language models~\cite{Mikolov2010-mp} based on LSTM  have been used to generate realistic text as well as being used in many downstream language understanding tasks~\cite{devlin-etal-2019-bert}. 
GANs~\cite{gan} are a recent type of generative model, which implicitly learn the underlying distribution through adversarial training. These have been successfully
used for image synthesis \cite{sngan, sagan, biggan}, audio synthesis 
\cite{wavegan, audiogan}, domain adaptation \cite{cyclegan, stackgan}, and other
applications \cite{xian2018feature, ledig2017photo}.

Our results unveil intriguing  connections between the learnability and complexity of sampling the output distribution through two generative model learning numerical experiments. 
In the first set of experiments,   we study how the learnability of various generative models for quantum sample learning task depend on  the complexity of quantum representation of the underlying probability distribution. Our experiments show that the required size of classical machine learning architectures grows exponentially with the number of parameters in a quantum circuit, which is determined by the number of qubits for a fixed-depth random circuit. As the qubit number exceeds a threshold, generative models on existing hardware platform failed to learn the output distribution of a deep random quantum circuit. For a fixed number of qubits, as circuit depth decreases below a threshold, the generative problem becomes   efficiently learnable.  Secondly, we unveil a similar correspondence between learnability and complexity with classically generated distributions that are close counterparts of random quantum circuits. We define related families of classical distributions using the fact that the bitstring probabilities of a random quantum circuit is close to Porter-Thomas distribution (see Definition 1 and \citet{boixo2018characterizing}). We study the learnability vs complexity of Porter-Thomas distributions by introducing structure through  re-ordering the bitstrings of a Porter-Thomas distribution from an ordered initialization. We can vary the  complexity of these orderings by introducing the $m$-bit subset parity function  
	\begin{align}\label{subsetparityEq}
     f_{sp}(x)= \text{mod}( x \cdot y, 2),
 \end{align}
 over bitstring $x$ (see Definition 6 in the following section) to determine the bitstring order. Note that some commonly used deep learning algorithms fail at discriminating the subset parity problem~\cite{shalev2017failures}. As we increase the Hamming weight of y, $m = |y|$, the classical samples becomes harder to learn. Finally, if we reorder the classical distribution according to a fully random permutation over all $2^n$ bitstrings, the learning task is again not efficiently learnable. 
The complexity of representing the subset function is at most linear in $n$, while the complexity of the most efficient representation known to-date for both a random permutation over $n$-bit strings  and a deep $n$-qubit random quantum circuit classically are  exponential in $n$. Our results therefore point to an unknown relation  between the learnability of bitstring samples and the underlying computational complexity of the distribution. 

We introduce the detailed problem definition and basic notations in Sec.~\ref{probdefSec}; discuss four generative models studied in this work in Sec.~\ref{GenModelsSec}, present the experimental results of generative model training on quantum sample learning in Sec.~\ref{ExpSec}, and discuss the connection between learnability and complexity in Sec.~\ref{LearnComplexSec}.

\section{Problem Definition }\label{probdefSec}
	Quantum mechanics permits the state of a quantum computation  on $n$ quantum bits~(qubits) to be in a superposition of  all $2^n$ states $\{0,1\}^n$. This is represented by a  wavefunction  as $ \ket{\psi} = \sum_{j=0}^{2^n-1} \alpha_j \ket{j}$
where the summation represents the coherent superposition over the computational basis denoted by  $\ket{j}\in \{\ket{0,\ldots, 0}, \ket{1, 0, \ldots, 0}, \ldots, \ket{1,\ldots, 1} \}$ with a complex number $\alpha_j$ as the associated amplitude. 
The quantum circuit transforms an initial state of quantum computer, say $\ket{0}$ to a final state $\ket{\psi}$ through a depth-$d$ quantum circuit which realizes the unitary transformation $U $. 
It is defined as a product of  single-qubit gate transformations $\{U_{k_i}\}$ and two-qubit gate transformations  $\{U_{k_i,k_j}\}$ as:
$
  U =\prod_{k=1}^d\left(\prod_{i=1}^n U_{k_i}\right)\left(\prod_{i=1}^n\prod_{j=1}^nU_{k_i,k_j}\right).
$ At each clock cycle $k$ only one non-identity gate can act on each qubit.
The quantum computation ends by applying a quantum measurement which outputs a computational basis state $\ket{j}$. The output probability distribution  $P(j)$ is determined by the quantum circuit $U$, acting on the initial state $\ket 0$, through Born's rule as
$  P(j) = \vert  U_{j,0}|^2. $
Notice that $U$ is a $2^n \times 2^n$ unitary. As the depth of quantum circuit $d$ and number of qubit $n$ gets bigger, the amount of classical computation required to sample from the distribution $P(j)$ grows exponentially.  At around $n\approx 32$,  the amount of memory needed to store all the amplitudes of each computational basis  exceeds the RAM space of an average laptop. And quickly, we lose any ability to calculate and thus understand the properties of these quantum distributions. However, given only samples from $P(j)$, we might still learn through generative models to reconstruct and thus characterize the underlying quantum distribution $P(j)$. Towards this goal, we define a new type of computational task, namely sample learning~(SL), as follows.

\noindent\textbf{Definition 1.} \textit{Sample learning}: given a set of $n$-bitstrings  $\mathbb{Q}$ of size $N_0$ drawn from the probability distribution $P(j)$, with $N_0\ll 2^n$,  learn to sample bitstrings strings with probability $P'(j)$ with statistical distance $\epsilon$: $\sum_j |P'(j) - P(j)| \le \epsilon $.\label{definition1ref}

Notice that we require the sample size  $N_0$  given to the generative model to be much smaller than the Hilbert space dimension. This is necessary not only to make the learning task not trivially learnable, but also to permit a fixed bound on the training time while scaling up the size of network architecture as $n$ increases. During training   we need a metric to evaluate the performance of the learning agent with  a set of bitstrings $\mathbb{G}$ produced by a learning algorithm. We use the linear cross entropy~(XEB) sample fidelity~\cite{boixo2018characterizing,supremacy2019quantum, bouland2019complexity,aaronson2016complexity,aaronson2019classical}
\begin{align}\label{eq:fidelity}
     F(\mathbb{G}) =2 \sum_{j \in\mathbb{G}} P(j) -1
\end{align}
The higher the fidelity is, the more truthful the samples are to the ones from the ideal distribution, which reaches maximum value  $F(\mathbb{G})=1$ for the deep random circuits in~\cite{supremacy2019quantum}. We also tested numerically that learning with this cost function achieves sampling that passes the chi-squared test~(see Appendix J).

\begin{figure}[ht]
\begin{center}
\includegraphics[width=0.5\columnwidth,trim=0 0 220 0, clip]{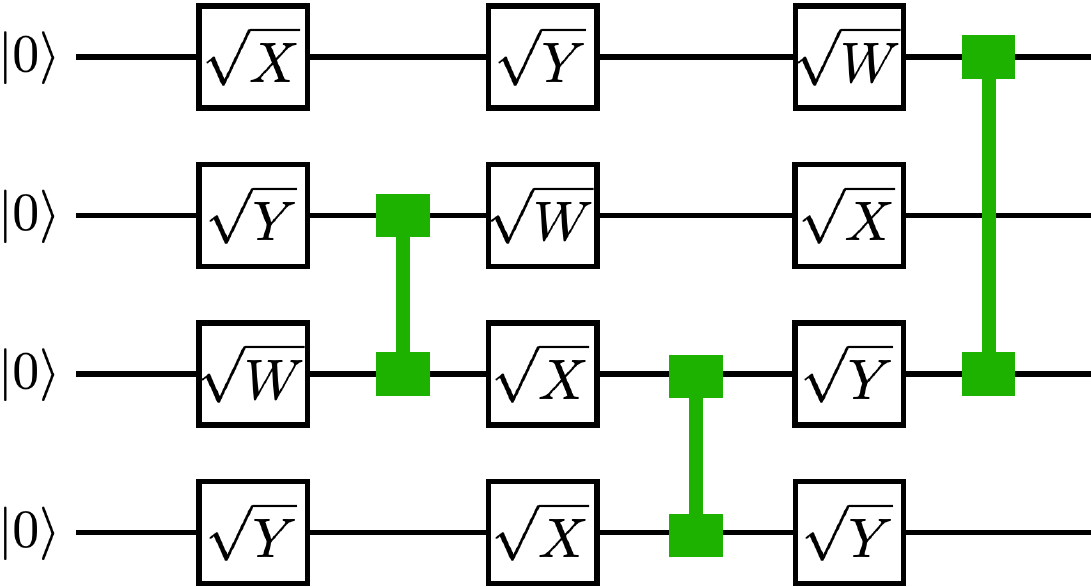}
\caption{A depth 2 random circuit, with $\sqrt{X}$, $\sqrt{Y}$ and $\sqrt{W}$ representing the single-qubit unitary, and the green line connecting two qubits representing a two-qubit CZ gate. All the gates are defined in    Table~1 in Appendix~I.\vspace{-20pt}
\label{depth4circuit}}
\end{center}
\end{figure}

To define quantum sample learning for random quantum circuit, we review the circuits used in the  ``quantum supremacy" experiments~\cite{supremacy2019quantum}, which consists of single-qubit quantum gates~(which apply a   unitary transformation on a single qubit), and two-qubit gates~(which apply a unitary transformation on a pair of qubits) chosen according to a specific pattern to maximally entangle all the bits through quantum correlations, see Fig.~\ref{depth4circuit} for a depth 2 random quantum circuit on four qubits. 

Approximately sampling from such random quantum circuits can be done efficiently with a quantum computer, but can take  exponential run time in the number of qubits using a classical computer.
It is shown in \citet{boixo2018characterizing} that the output of a sufficiently deep random quantum circuit defined above approximates that of a Porter-Thomas distribution~\cite{PoterThomas1956}. More specifically, under the ideal Porter-Thomas distribution  the probability that a $n$-bit string $z$ with probability in $[p, p + dp]$ is sampled equals: 
\begin{align}\label{PTdistEq}
    p2^{2n} e^{-2^n p} dp.
\end{align}
 Plugging this into the fidelity expression (Eq.~(\ref{eq:fidelity})),   gives us $ F(\mathbb{G}) =1$ for $N\gg 1$.  Depending on how the given quantum samples are produced, we have two categories of quantum SL problems defined as follows.
 
\noindent\textbf{Definition 2.} Experimental quantum sample learning (QSL): SL from Definition 1, with the given training quantum sample $\mathbb{Q}$ generated by an experiment, which includes noise and errors in its output. 

\noindent\textbf{Definition 3.} Theoretical QSL: SL  with the given training quantum sample $\mathbb{Q}$ generated by a ideal  simulated quantum computation.

QSL is the quantum counterpart of the image generation problem, with pixel values replaced by bit string values. And   experimental QSL can then be compared to unsupervised image denoising,  which permits additional noise caused by imperfections in quantum computers' actuations.   We study both types of QSLs in this work utilizing quantum bit strings obtained from:  running random quantum circuits on a laboratory quantum computer for experimental QSL task; and   simulation of sampling from perfect quantum distributions for theoretical QSL tasks~(see Appendix for detailed simulation methods). Both types of data and training code for QSL has been released publicly.  

In order to compare the efficiency of sample learning with related classical distributions of variable complexities, we introduce the following definitions.

\noindent\textbf{Definition 4.} \textit{Porter-Thomas Distribution over $n$ bits}:  the portion of the bitstrings that have probabilities in the range $[p, p+ dp]$ decays exponentially as Eq.~(\ref{PTdistEq}).

\noindent\textbf{Definition 5.} \textit{Ordered Porter-Thomas Distribution over $n$ bits}: as in Definition 4, but in addition the  bitstrings are ordered according to the integer value represented by each bitstring. That is, the probability of each bitstring decreases as its integer value represented by the bitstring increases.

It is not hard to see that ordered Porter-Thomas has an efficient classical representation of an exponentially decreasing probability density. Now, we add a complexity to the ordered Porter-Thomas by adding the subset parity determined reordering defined as follows.

\noindent\textbf{Definition 6.}\label{Def6} \textit{$m$-bit Subset Parity ordered Porter-Thomas Distribution over $n$ bits} is defined by reordering the distribution of Definition 5. Given a random $n$-bitstring $y$ with $m$ nonzero entries, the subset parity of bitstring $k$ is defined by $\mod(k \cdot y, 2)$. In the $m$-bit subset parity order,  bitstrings with even (0) subset parity appear before bitstrings with odd subset parity, while preserving the original order within the subset of the bitstrings sharing the same subset parity value. 

It is also straightforward to see that the representation of subset parity ordered Porter-Thomas is at most linear in $n$ bounded by the length $n$-bit binary vector $y$. In contrast, we  can define another classical distribution below that  has an exponential in $n$ number of parameters in its representation.

\noindent\textbf{Definition 6.} \textit{Random permutation Porter-Thomas Distribution over $n$ bits}: the distribution of probabilities is Porter-Thomas  as in Definition 4, but the ordering of the bitstrings is given by a random permutation over all bitstrings from an initially ordered value.

We can now define two additional learning tasks which we benchmark against the quantum distribution to understand the intimate relation between complexity and learnability. 

\noindent\textbf{Definition 8.} \textit{Subset Parity sample learning}: Given a set of $n$-bit strings $\mathbb{S}=\{j \}\in \{0, 1\}^n $  of size $N_0$, with $N_0\ll 2^n$, drawn from the probability distribution $P(j)$ defined in  Definition 6, generate new samples satisfying criteria in Definition 1.

\noindent\textbf{Definition 9.} \textit{Random Porter-Thomas sample learning}: Given a set of $n$-bit strings $\mathbb{S}=\{j \}\in \{0, 1\}^n $  of size $N_0$, with $N_0\ll 2^n$, drawn from the probability distribution $P(j)$ defined in  Definition 7, generate new samples satisfying criteria given by Definition 1.

\subsection*{Related Works}
Using machine learning tools to more efficiently sample from quantum systems has been explored using the models of restricted and deep Boltzman machines and quantum Monte Carlo methods~\citet{carleo2018constructing, PhysRevB.61.R16291, PhysRevB.95.035105}. Generative models for learning bitstrings produced by classical circuits is discussed in \citet{sweke2020quantum}, where an exponential separation between classical learning models and a tailored quantum learner is proven based on cryptographic conjectures.
However, no prior work exists to provide either  heuristic or theoretical analysis on the learnability and complexity of QSL architectures  in the context of random quantum circuits.

\section{Generative Models }\label{GenModelsSec}

In this section, we introduce four kinds of generative models covering two types of learning objectives: explicit methods based on maximum likelihood optimization, and implicit distribution learning through GANs; as well as two types of architectures: one-shot vs autoregressive sequential generation. Autoregressive architectures involves the use of neural networks that generate the next bit of a sample based on the previous sampled bits whereas one-shot generation generates an entire bitstring at once.  
\subsection{Deep Boltzman Machine}

Deep Boltzman machine~(DBM) \cite{salakhutdinov2009deep} is a type of generative model which explicitly parameterizes the dependency of the probability distribution through the use of a visible layer~(representing the actual sample values) and deep hidden layers, such that arbitrary non-local dependencies between the visible units can be indirectly realized through its connection with common deep layer units. We give a direct mapping form a quantum circuit to a DBM in Appendix.~\ref{app:dbm}. We also show that it has an exponential number of parameters in the depth of quantum circuit and the number of qubits. 




\subsection{  Generative Adversarial Networks}\label{GANSec}
Generative adversarial  networks~(GANs) \cite{gan} are special instances of  generative models, which  learn the underlying data distribution  $p_{data}(x)$ given a training data set $\{x_i\}$. This is done by training a generator model $G$ to generate realistic samples while a discriminator model $D$ is trained to discriminate between samples from $G$ and the training set. The generator improves by fooling the discriminator since its weights are updated using gradients that flow through the discriminator. After their formulation, GANs have quickly become one of the most popular  generative models. 
GANs have become popular due to two factors: 
First, they tend to produce samples that -- at least in isolation -- are 
difficult to distinguish from training data (this is built into their 
objective function).
Second, because their loss function is implicit, it is simple to modify them
for use in a new domain or application.
It generates samples from a single step through a neural network  as apposed to   commonly used Markov chain method for FVBN or Boltzmann machines which takes many steps of sampling to realize a stochastic approximation.  Since GANs have been used to successfully model complex distributions of high modalities
\citep{CRGAN,openquestions,ylg,smallgan,progan,skillrating,biggan,sagan}, they are promising candidates for QSLs.

\subsubsection{One-shot GAN}
To train GANs to generate sample bit-strings with $n$ qubits, we used a fully connected
generator and discriminator with a uniform prior.
The generator outputs floating point numbers between 0 and 1 that are eventually
rounded to the nearest integer to create proper bit-strings.
In order to prevent the discriminator from easily distinguishing between `real'
and `fake' data early on in training, we had to add noise to the inputs to the 
discriminator as is commonly done \cite{improvedtechniques}.
For small numbers of qubits, this technique performs reasonably well,
as can be seen in Figure \ref{summaryplotQSL} and \ref{theoreticalsummaryplotQSL}. For larger numbers of qubits, we found that the distribution implied by the generator
had support mostly overlapping the distribution implied by the circuit, but bit-strings
were mis-weighted (that is, the GAN outputs most of the right bit-strings, but in the
wrong proportions).
\citet{DRS}  designed a GAN to fix this problem, so we used it and found that it
did indeed help, though the performance as measured in fidelity is still lower
than the theoretical bound and the performance achieved by the language model
(see Section \ref{LMSec}). The choice and optimization of hyperparameters, and detailed implementations are described in Appendix~C.

\subsubsection{Autoregressive GAN}\label{SecGANSec}
As opposed to standard GANs which have typically been used to generate images and other non-sequential data, autoregressive GANs have been used to generate text as an alternative to maximum likelihood models. As with GANs, this approach has the advantage that no explicit loss function for the generator needs to be specified. The discriminator also inspects the entire generated sentence at once and so its gradients can improve the generator in a holistic way. Some common approaches include SeqGAN \cite{seqgan} and MaskGAN \cite{maskgan} which have been used to generate fake movie reviews and news articles. We use the   open-source implementation of SeqGAN~\cite{seqgan} with the LSTM generator for  QSL, see Appendix D.

\subsection{ Autoregressive Maximum-Likelihood Models }\label{LMSec}
Recurrent and autoregressive models such as long short-term memory models (LSTMs) \cite{Hochreiter1997-mo} are commonly used to model and generate sequence data. Recurrent neural nets are models that consist of a hidden memory that stores a representation of the data observed so far. The weights of the model are shared across time making them ideal to sequence and especially variable-length data. LSTMs are a variant of recurrent models that add an input gate, an output gate and a forget gate, where these gates are able to regulate what information flows into and out of the hidden memory. This enables them to learn more complex sequence data.

Language modeling \cite{Mikolov2010-mp} is a common use of these models where the model predicts a probability distribution over the next input given the previous inputs so far. Typically these inputs are words, byte sequences or individual bytes. In general, autoregressive models are expected to perform better with sequences since it's easier to generate a sentence sequentially (word-by-word) than the entire sentence at once. In our case, the inputs to the language model are samples of bitstrings and the model is trained to predict the next bit given the bits observed so far, starting with a start of sequence token. We use a standard LSTM language model with a logistic output layer. To sample from the model, we input the start of sequence token, sample from the output distribution then input the result as the next timestep. This is repeated until the required number of samples is obtained.  We detail the choice and optimization of hyperparameters, and experimental implementations are described in Appendix~E.

\section{Experiments}\label{ExpSec}
We trained three model architectures on the task of generating bit strings (QSL) and measure the fidelity as in Eq.~\eqref{eq:fidelity} of bitstrings sampled from the model. The training set consists of 500000 samples from the corresponding theoretical or experimentally obtained distributions. The supplementary material describes how those distributions were obtained~\cite{open_source_code}. Various sizes of models (256--2048 units) were trained to measure the model capacity needed for the quantum distributions. The model hyperparameters are listed in the supplementary material. Fidelity was calculated with 200000 samples from each of the models.

We evaluate the models on the QSL task for every two qubit size between 12 and 24 to understand how the models performance scale with the number of qubits. Additionally, we evaluate the models on QSL for 24 qubits with each circuit depth from 2 to 12 to understand how performance scales with depth. The Hilbert space dimension for the 12 to 24 qubits ranges from 4,096 to 16,777,216.

\subsection{Architecture Comparison} 
\begin{figure}[ht]
\begin{center}
\includegraphics[width=1\linewidth]{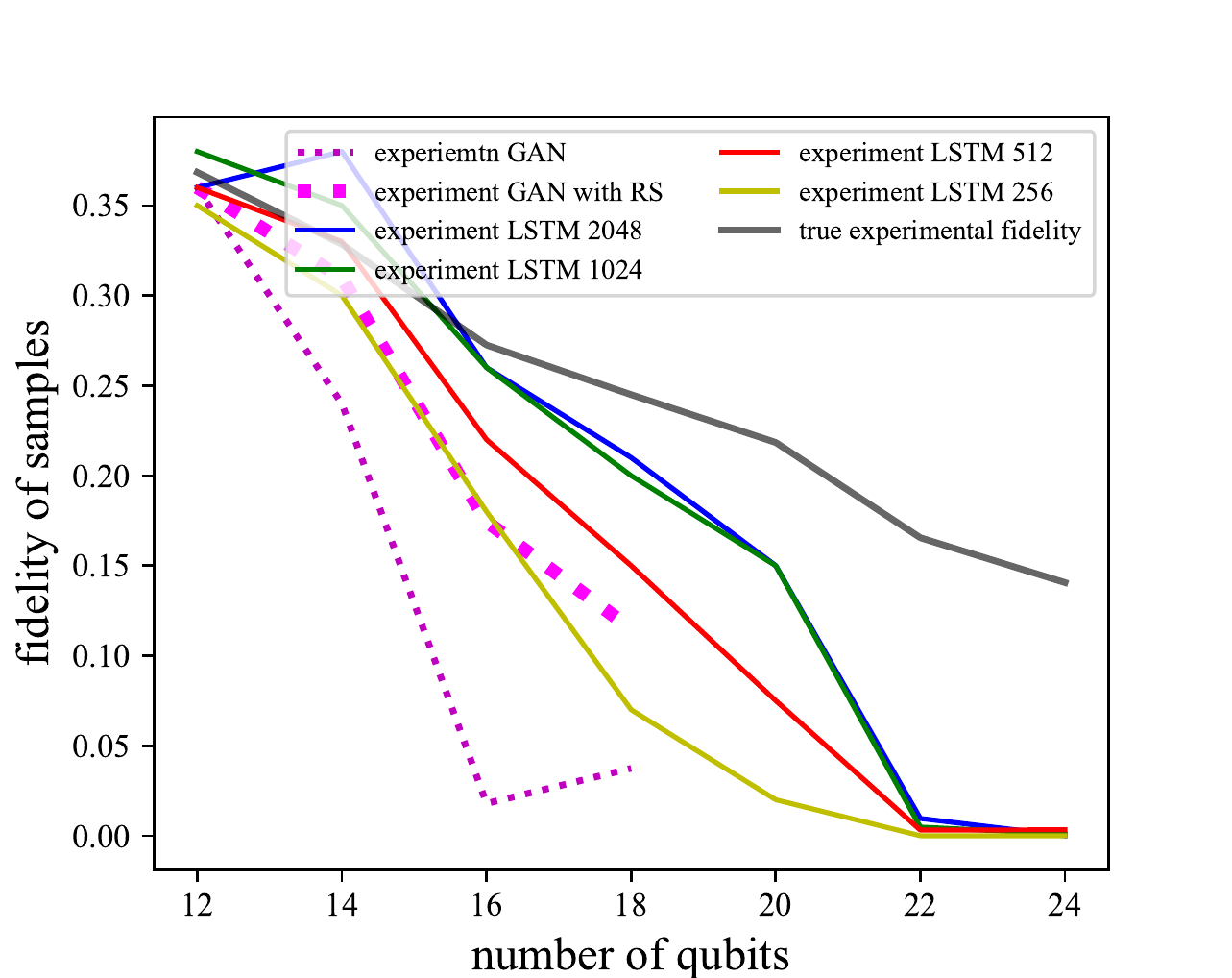}
\caption{Performance of GAN, rejection sampling based GAN and LSTM for learning quantum distribution from    experimentally measured quantum circuit outcomes~\cite{supremacy2019quantum}.
\label{summaryplotQSL}}
\end{center}
\end{figure}

The performance of various architectures described in the previous section for solving experimental QSL  tasks for 12 to 24 qubits is summarized in Fig.~\ref{summaryplotQSL}. Notice that the true experimental bitstrings are given fidelity less than one due to the noisy nature of experimental implementation. A faithful learning should achieves a fidelity that equals the experimental fidelity. However, it is possible that the learned model is not capturing such underlying realistic noise in the samples, and achieves higher fidelity than actual experimental fidelity, see for example blue and green curves representing the performance of LSTM with 1024 and 2048 hidden variables per layer.

Of the models we've considered, the standard one-shot GAN model performed poorly. This GAN only succeeds in learning the experimental QSL for 12 qubits while requiring a range of 1 million to 17 million  training parameters. Rejection sampling is able to improve upon the original GAN by a factor of five in cross-validation fidelity, but is still unable to reach a fidelity comparable with the true experimental fidelity. The LSTM models are able to model the experimental fidelity with larger LSTMs up to around 18 qubits after which the performance falls off. The results show that larger LSTMs are needed for higher fidelity samples, especially from 16 qubits where the LSTM of 256 hidden units achieves 0.15 fidelity and the LSTM with  1024 hidden units achieves 0.27. This is intuitive since the Hilbert space grows exponentially with the number of qubits.

\begin{figure}[h!]
\begin{center}
\includegraphics[width=1\linewidth]{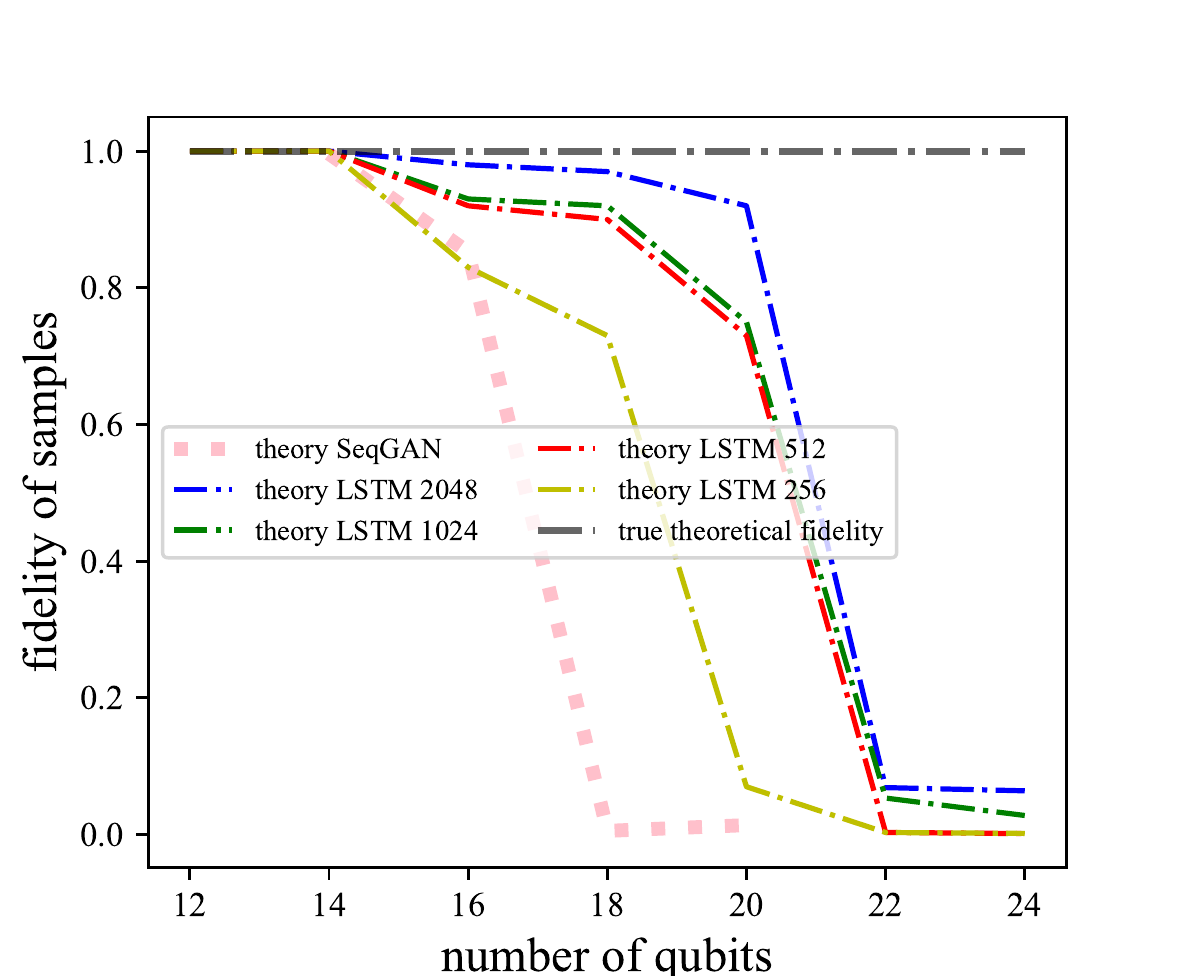}
\caption{Performance of  rejection sampling based GAN and LSTM for learning quantum distribution from  theoretically generated bitstrings from quantum circuit simulation.
\label{theoreticalsummaryplotQSL}}
\end{center}
\end{figure}

For the results with theoretical quantum distributions in Fig.~\ref{theoreticalsummaryplotQSL}, the autoregressive GAN (SeqGAN), is able to learn the noiseless theoretical distribution up to 16 qubits. In this case, the GAN models are outperformed by LSTMs using maximum likelihood optimization. The plot shows the general pattern that higher capacity LSTMs (more hidden units) are able to model larger numbers of qubits with higher fidelity. LSTMs with 2048 units are able to learn up to 20 qubits well but fall off in performance after that. This may be due to the increasingly high capacity requirements needed to learn longer qubit lengths.

The improved performance of the autoregressive models (LSTMs and SeqGAN) relative to the one-shot GAN model indicates that there is hidden sequential structure in the random quantum circuit output for both shallow and deep depths. This is surprising given the theoretical proof that at the maximum circuit depth the quantum distribution is close to that of a Porter-Thomas distribution over randomly ordered bit strings~\cite{boixo2018characterizing}. The advantage in generative models that explicitly utilizes sequential structures in their networks for learning QSL  unveils the  inherent sequential nature of the quantum distribution at the depths studied here. 

 In addition to XEB fidelity, we also performed the $\chi^2$ test to the generated samples, which is a more robust and stringent measure than XEB.  But  the two measures are consistent with each other given enough training time for the generative models. We discuss in more detail in Appendix.~I, the definition of $\chi^2$ test and detailed empirical distribution histogram comparison between learned bitstrings and expected bitstring samples. 
 
\subsection{Model Complexity Scaling }
\begin{figure}[ht]
\begin{center}
\includegraphics[width=1\linewidth]{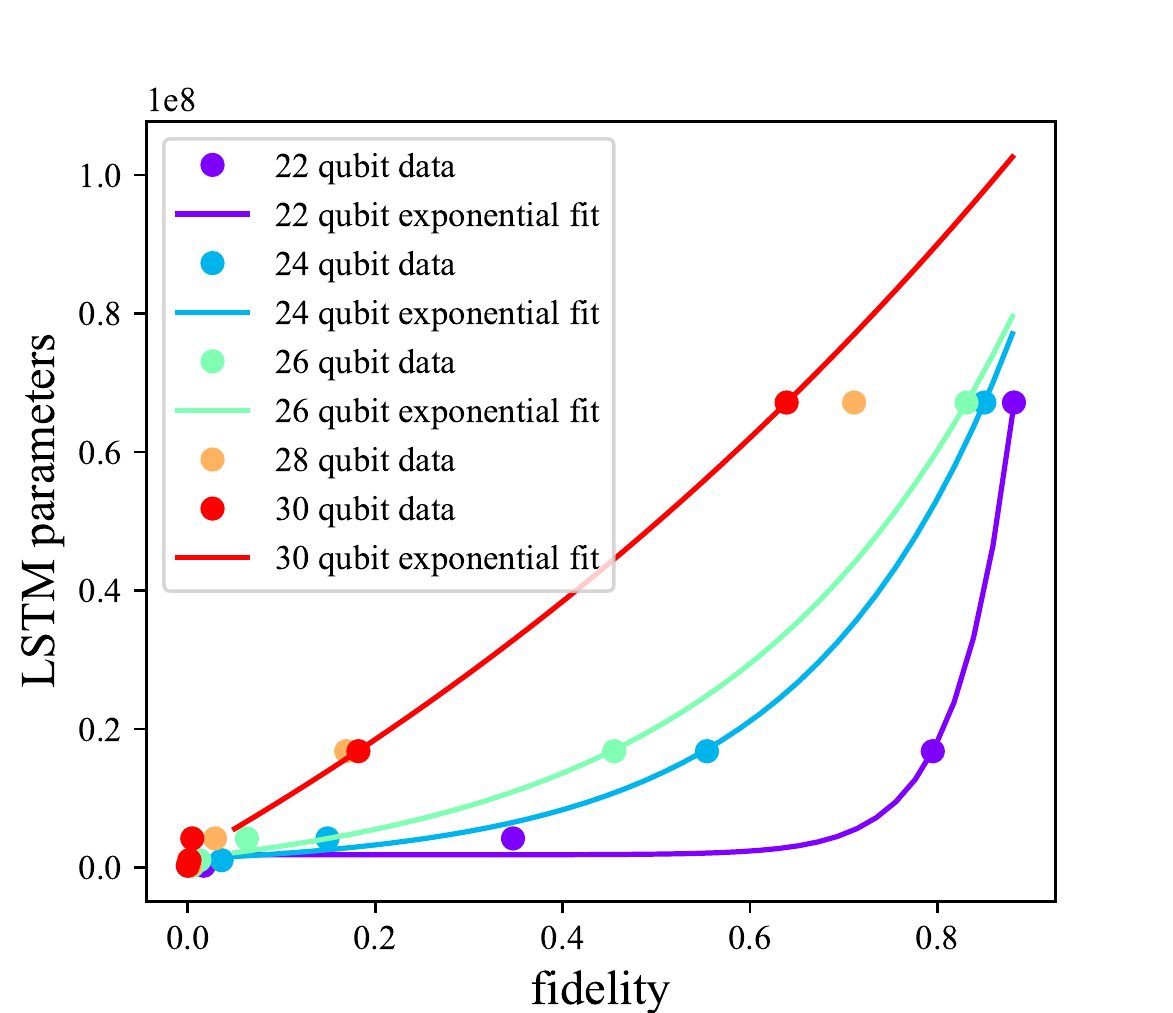}
\caption{The data (dot) and the fitted exponential function $y = a \cdot \exp(b \cdot x) + c$ (solid line) for the required number of parameters in LSTM as a function of number of qubits for achieving different levels of  fidelity.
\vspace{-20pt}
\label{lstmsizevsqubitnumber}}
\end{center}
\end{figure}

\begin{figure}[htb]
\begin{center}
\includegraphics[width=1\linewidth]{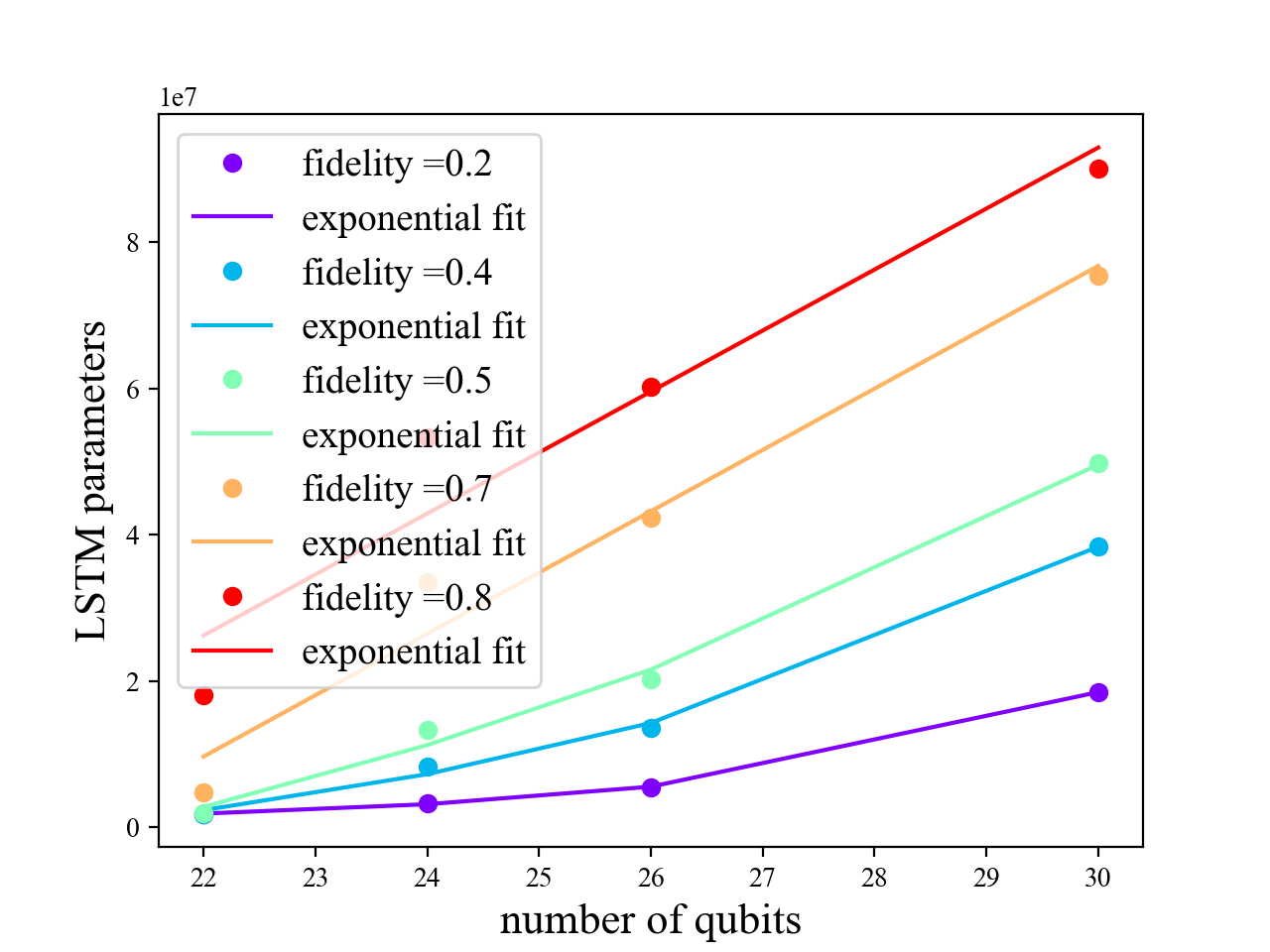}
\caption{The minimum  number of LSTM parameters as a function of the number of qubits for achieving different levels of  fidelity based on the fitted function from 22, 24, 26, and 30 qubit data (dots), and exponential function $y = c \cdot \exp(d \cdot x) + f$ (solid line) .
\label{lstmvsnumbqubit}}
\end{center}
\end{figure}

 In order to study the relationship between generative model capacity and the QSL problem size, we focus on LSTM architecture which is the best in regard to model efficiency and learning accuracy. We train and evaluate LSTMs of varying sizes under 800 training epochs for varying numbers of qubits. The achieved fidelity as a function of the total number of LSTM parameters is shown in Fig.~\ref{lstmsizevsqubitnumber}. The total number of LSTM parameter is determined by the number of hidden units of LSTM by a nonlinear relation. Fig.~\ref{lstmsizevsqubitnumber} covers LSTMs of 256 hidden units to 4096 hidden units. Remarkably, the data fits an exponential function for all data except the 28 qubit experiment. With this exponential function obtained from numerical experiments, we can infer the required LSTM size as a function of the number of qubits as show in Fig.~\ref{lstmvsnumbqubit}.
 The total number of parameters that specify a depth-$d$ random circuit of $n$ qubit is roughly $1.5 \times d \times n$. For quantum supremacy experiment, we have $d=14$, so the number of parameters of roughly $21 \times$ the number of qubits. Therefore, the scaling of required LSTM parameters as a function of the quantum circuit parameters is also exponential, as in Fig.~\ref{lstmvsnumbqubit}. In stark contrast, bitstrings generated through classical means can be efficiently represented by LSTM, as is shown in 
 \citet{capacity},  who give evidence that LSTMs can store a fixed amount of classical 
 information (5 bits) per parameter.  
 We also find that as we increase the QSL problem size to 34 qubits, all the generative models failed to learn the theoretical QSL due to the inefficiency of sampling from a theoretical distribution over $2^{34}$ bitstrings. And they also fail at the experimental QSL learning task for 34 qubits with a maximal achievable fidelity below $0.001$ given the true experimental fidelity for 34 qubits is around $0.04$.
 
\section{Learnability and Complexity}\label{LearnComplexSec}
In this section, we delve deeper into possible mechanisms that defines   the learnability of given bitstring samples produced both quantum mechanically and classically.
From the previous section, we show that the required classical neural  network complexity scales exponentially with the number of qubits. Eventually the memory requirements  exceeds the permissible values as qubit number increases, and the QSL becomes un-learnable. Can we attribute such transition in learnability to the complexity of the classical representation of the underlying probability distribution?  This conjecture can be tested if given a nob to tune the complexity of the underlying probability distribution, and evaluate the corresponding learnability of the associated samples. We perform such test with both quantum and classical samples.

\begin{figure}[h!]
\begin{center}
\includegraphics[width=1.0\linewidth]{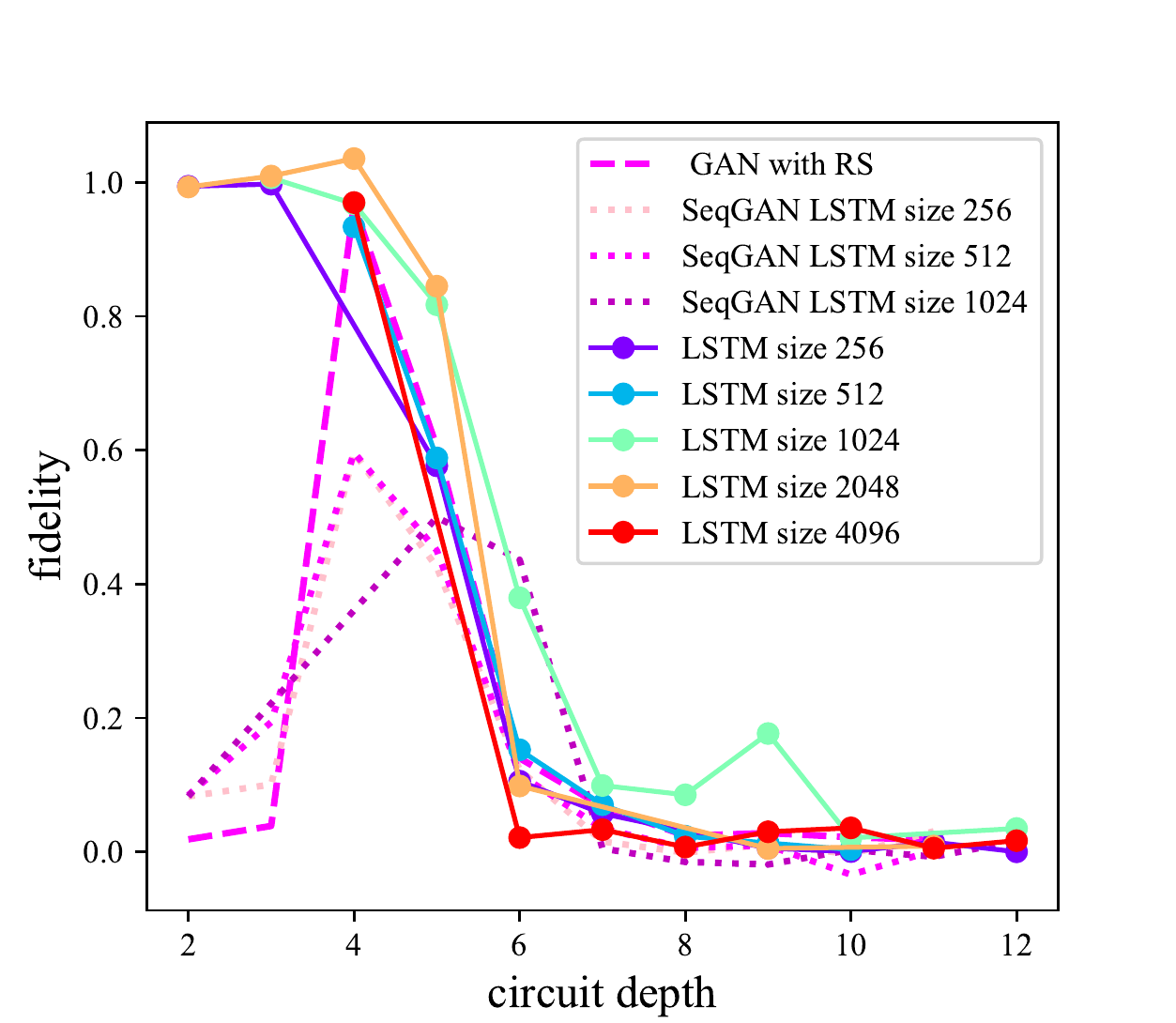}
\caption{The  fidelity of the learned samples from quantum circuit on 24 qubits of depth 2 to 12. A transition at around depth five occurs, beyond which the output of the quantum circuit becomes hard to learn.\vspace{-20pt}
\label{LSTM22-24results-400Epoch}}
\end{center}
\end{figure}

For quantum samples, two parameters dictate the complexity of the underlying  distribution: number of qubits and the depth of the circuit. As we increase the number of qubits,  as shown in Fig.~\ref{lstmvsnumbqubit}, the required classical representation scales exponentially. We can also fix the number of qubits but vary the circuit depth. This is realized by taking one layer of gates away from the random quantum circuit   at a time  and observing how learnability of the best  generative models varies. We  observe an intriguing transition at circuit depth 5, where LSTMs of any size studied~(ranging from smaller to larger than the Hilbert space of 24 qubits) fail  to learn with high enough  fidelity if given a short amount of training time that is comparable to other generative models under consideration~(GAN and SeqGan), see Fig.~\ref{LSTM22-24results-400Epoch}. This transition corroborates with the collapse of higher order dependence in the conditional probability distribution shown in Fig.~\ref{maximalconditionalProbfig} and the convergence to the Porter-Thomas distribution in entropy shown in Fig.~\ref{entropy20to26}. The collapse of these higher order dependencies imply that the sequential dependence between bits of the bitstrings become scrambled and thus evened out when depth increases above five regardless of the number of qubits~\cite{boixo2018characterizing}. With lower depths, even though the Hilbert space is of size 16,777,216 (comparable to the same number of parameters as a  LSTM with 2,048 hidden units), significantly smaller LSTMs (the LSTM with 256 hidden units  only has 264,449 parameters) can still learn to generate bitstring samples at a high fidelity. This implies that there remains significant sequential correlations in the bitstrings at lower depths that permits an efficient classical representation of its distribution.


 If the learnability-complexity correspondence hold in QSL, it should also pass similar test in classically generated samples. Indeed, the classical counterpart of deep random quantum circuit output is a Porter-Thomas over bitstrings with randomly permuted order. Since to represent this distribution, we need to specify the order of each one of $2^n$ bit strings, which in the generic case does not have an efficient lower dimensional representation, and thus high in the complexity measure. We then chose a well-studied functional class, $m$-bit subset parity, to re-order the bitstring orders of an ordered Porter-Thomas distribution, and $m$ serves as a nob that tunes the complexity of this classically generated distribution. Such subset parity reordered distribution is formally defined in Definition 6 of Sec.~\ref{Def6}.

\begin{figure}[h!]
\begin{center}
\includegraphics[width=1.1\linewidth]{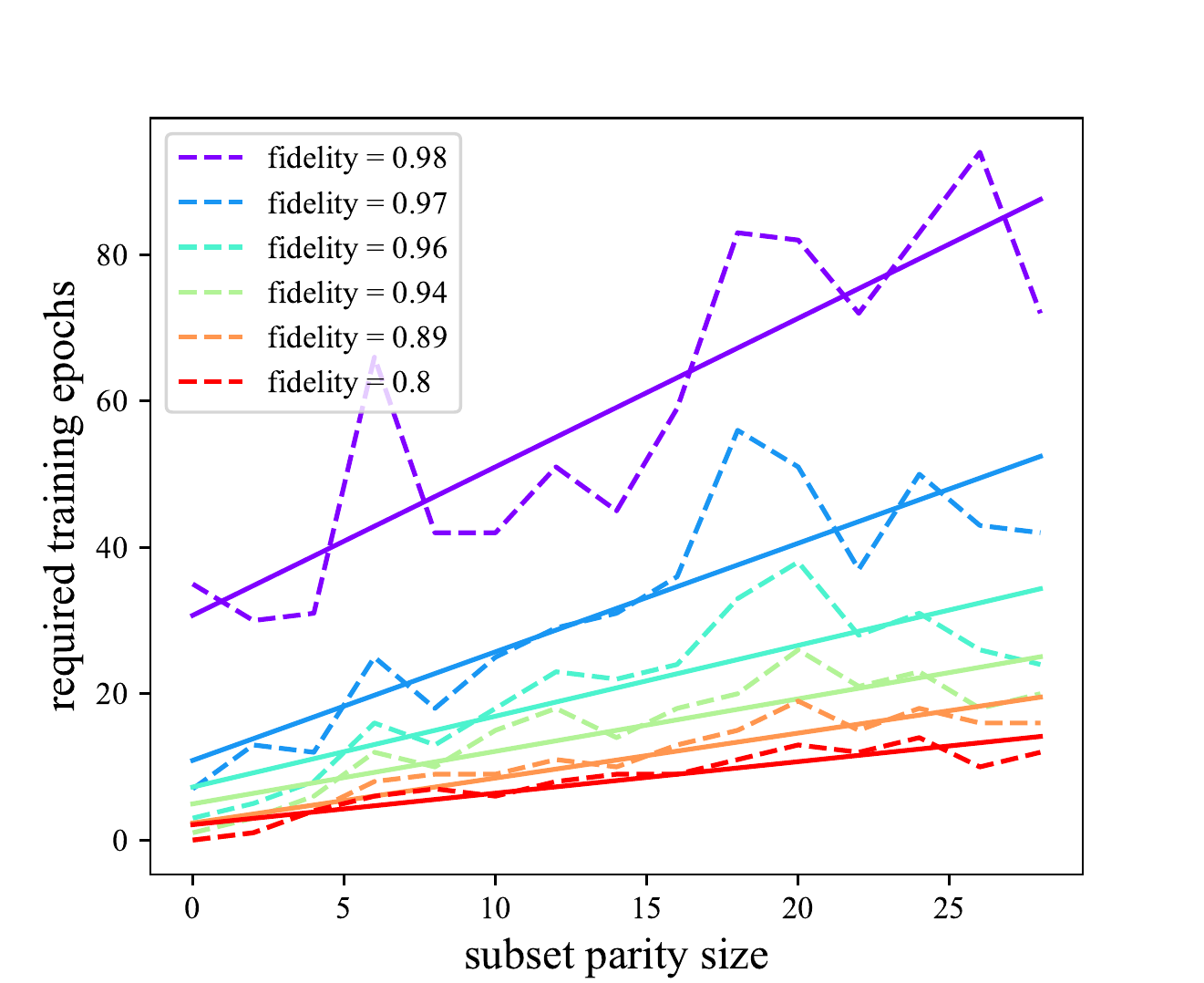}
\caption{Dotted lines: required training epochs to achieve a given fidelity as function of $m$ for $m$-bit subset parity ordered Porter-Thomas learning. Solid line represents a linear fit, dashed line represents the data of training LSTM of size 256.
\label{28qubit-subsetparity}}
\end{center}
\end{figure}

 We show with our experimental learning of this two classes of samples with the most effective geenerative model, LSTM, that Porter-Thomas with random permutation is as hard as if not harder than quantum distribution of the same size. But once we introduce subset parity ordering to the Porter-Thomas distribution as in Definition 6, the LSTM model performs well and achieves $0.98$ in merely $80$ epochs for the largest subset (28)bits, see Fig.~\ref{28qubit-subsetparity}. Although it is proven~\cite{shalev2017failures} that the subset parity is notoriously hard for gradient based method, which is used in LSTM update, since the gradient of an $m$-bit subset parity vanishes exponentially in $m$, an LSTM of 256 hidden variables is sufficient to achieve close to identity fidelity fairly quickly. Moreover, the learning efficiency decreases with $m$, while remaining learnable, see Fig.~\ref{28qubit-subsetparity}.  In contrast, the same architecture failed to learn a randomly ordered Porter-Thomas distribution, see Appendix H.
The only difference between these two drastically different performances is how we re-order the Porter-Thomas distribution: for  $m$-bit subset parity learning the order is specified by a random $n$-bit binary vector, for randomly permuted Porter-Thomas, the order is specified by permutation matrix with $O(2^n)$ independent elements.
Finally we also notice that  the quantum distribution generated by a deep random quantum circuit, albeit possesses an efficient quantum representation in the form of a quantum circuit lacks polynomial in qubit number classical representation, is as hard to learn as a randomly permuted Porter-Thomas distribution. 
These evidences corroborates   the conjectured correspondence   between learnability and complexity: distributions that possesses an efficient~(polynomial in $n$ many parameters) classical representation can be efficiently learned with generative models.  This is the first direct evidence of the exponential separation in the sample learning complexity of a distribution generated by quantum circuit  from that generated by classical circuits.

\section{Conclusion}

 

In this work, we apply four kinds of most widely used generative models to the task of quantum sample learning, to unveil intriguing  properties about the quantum distribution generated from random quantum circuits otherwise inaccessible through traditional top-down analysis. First, we show that the model capacity must increase exponentially in the number of qubits  to achieve a fixed fidelity. We show this for an explicit DBM model and also experimentally for the deep learning models.  Secondly, we discover a persistent sequential dependence between bit values in both shallow and deeper random quantum circuit output distributions from \cite{supremacy2019quantum} as evidenced by the superior performance of the autoregressive models compared to GANs. We also show that (similarly to text) maximum-likelihood trained models are able to generate samples closer to the quantum distribution than GAN models, which typically suffer from mode dropping.  
Finally, we experimentally verify the conjectured  learnability and complexity correspondence in learning generative models. We train LSTMs against both quantum samples and classical samples of varying degrees of complexities in their underlying probability distributions. For quantum samples, we  observe a transition in circuit depth below which a machine learning agent is able to learn quantum distribution using significantly less than Hilbert space amount of training parameters, which approximately coincides with the depth below which  an efficient classical approximation exists~\cite{napp2019efficient}. For classical samples, we show that by enforcing a classical structure, $m$-bit subset parity, onto the Porter-Thomas distribution that is the classical counterpart of quantum distribution of random circuits, the sample learning task in Definition 1 becomes efficiently learnable. The training time scales with the complexity of the classical representation of the underlying distribution.

These findings  provide new insights into the complexity of quantum distributions prepared by quantum computers, and point to the interesting regime of low-depth quantum circuits, which already demonstrates similar sequential structure to that shown in language. Additionally, the new QSL task can be used as a benchmark for generative models, where the complexity and difficulty of the task can be easily controlled and models can be evaluated exactly since the true distribution is known.

\vspace{5mm} 

Open source code of this paper is available in GitHub repository~\cite{open_source_code}.

\begin{acknowledgments}
We like to thank useful feedback from Dave Bacon, John Platt and Alexander Zlokappa on the manuscript. Quantum simulation and model training in this paper were implemented using Cirq~\cite{cirq} and TensorFlow~\cite{abadi2016tensorflow}.
\end{acknowledgments}
 
\clearpage
\newpage
\onecolumngrid
\appendix
\title{Appendix for: Are Quantum Samples Classically Learnable?}
\section{Sampling with Random Quantum Circuit }

In this section, we provide details on the type of probability distribution  used for demonstrating the quantum speedup: sampling   from random quantum circuits\cite{supremacy2019quantum}. The underlying distribution  $P(z)$ given by a random quantum   circuit realize  in  ~\cite{supremacy2019quantum} and also studied in this work is defined by the circuit  unitary transformation $U $ as:
\begin{align}
    P(j) = |\bra{j} U  \ket{0^n}|^2
\end{align} where we use $j$ to represent an $n$ bit string, and $\ket{j}$ as a computational basis state representing the same bit string. A depth $d$  unitary transformation is realized by $ d$ cycles of interlaced single and two qubit gates of forms:
\begin{align}
     U  = \left[\prod_{j=1}^d U_1(j)U_2(j)\right] ,
\end{align}  
where $U_1(j)=\prod_{i=1}^n U_{j_i}^1$ is a product of $n$ single qubit gates each randomly chosen from a given set $\mathbb{S}_1$ acting on each one of $n$ qubits. In existing proposals, $\mathbb{S}_1=\{\sqrt{X}, \sqrt{Y}, \sqrt{X+Y}\}$ each defined as:
\begin{align}\label{singleqGateset1}
  &  \sqrt{X} = e^{-i\frac{\pi}{4}\sigma^x}=\frac{1}{\sqrt{2}}\left[\begin{matrix} 1 & -i\\ -i & 1 \end{matrix}\right],\\
    &    \sqrt{Y} = e^{-i\frac{\pi}{4}\sigma^y}=\frac{1}{\sqrt{2}}\left[\begin{matrix} 1 & -1\\ -1 & 1 \end{matrix}\right],\\\label{singleqGateset3}
        & \sqrt{X+Y} = e^{-i\frac{\pi}{4}\frac{\sigma^x + \sigma^y}{\sqrt{2}}}=\frac{1}{\sqrt{2}}\left[\begin{matrix} 1 & -\sqrt{i}\\ -\sqrt{i} & 1 \end{matrix}\right].
\end{align}
And the two-qubit gate layer is defined as
\begin{align}
    U_2(j) =\prod_{i=1}^n U_{j_i}^2= \prod_{i,j \in \mathbb{N} } \text{fSim}(\theta,\phi)_{i,j}
\end{align} which is a product of two-qubit gate  $\text{\text{fSim}}(\theta,\phi)$ acting on the neigboring qubit $i,j$ chosen from the set of connected qubit pairs permited by the hardware architecture $\mathbb{N}$. The two qubit gate is parameterized by two rotation angles $\theta$ and $\phi$ as:
\begin{align}
    \text{fSim}(\theta,\phi) =
    \left(\begin{matrix}
    1 & 0 & 0 & 0 \\
    0 & \cos(\theta) & - i \sin(\theta) & 0\\
    - i \sin(\theta) & \cos(\theta) & 0 & 0\\
    0 &0 & 0& e^{-i \phi}
    \end{matrix}\right)
\end{align}
In another formulation of random circuit~\cite{shepherd2009temporally}, the two qubit gate is chosen to be CZ defined by 
\begin{align}
    \text{CZ}  =
    \left(\begin{matrix}
    1 & 0 & 0 & 0 \\
    0 & 1 & 0 & 0\\
    0 & 0 & 1 & 0\\
    0 &0 & 0& -1
    \end{matrix}\right).
\end{align} In this work we study the type of random circuit used in \cite{supremacy2019quantum}, which is has no fundamentally distinct properties from that with CZ gates. For the completeness of our study, we will specify the construction of all above mention gate units with Deep Boltzman Machine in the next section.

\section{Characteristics of Quantum Distribution}
Before exploring the hardness of  QSL tasks defined above, we first investigate the mathematical properties of the quantum distributions from random ``quantum supremacy" circuits in greater detail, in regard to their  entropy  and correlations. 
 
\subsubsection*{Entropy}

\begin{figure}[ht]
\begin{center}
\includegraphics[width=0.6\linewidth]{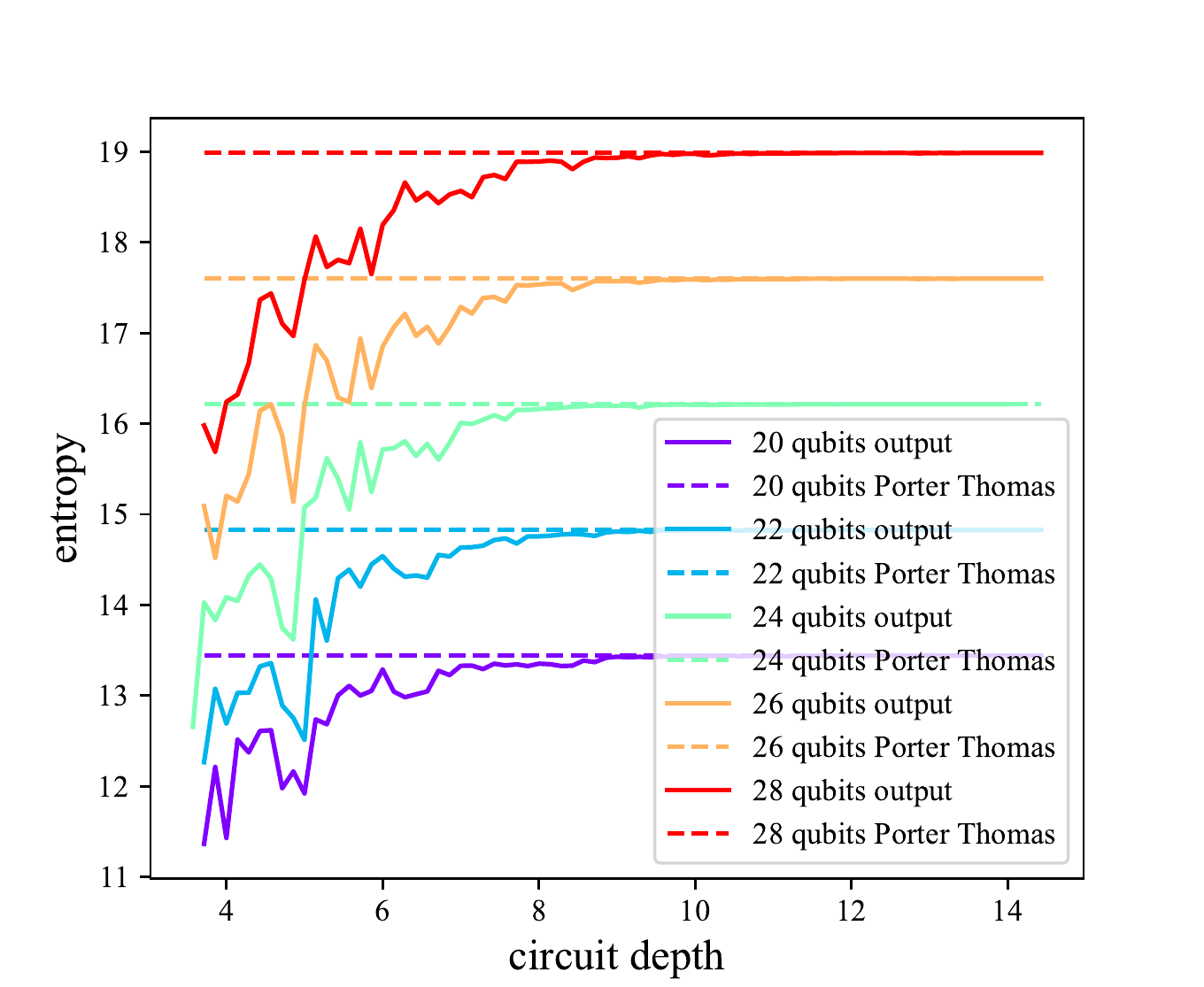}
\caption{The entropies of quantum distributions given by 20 to 28 qubits  random quantum circuits  of depths ranging from $4$ to $14$~(solid), compared against the entropy for the bit strings draw from perfect Porter-Thomas distribution~(dashed) defined in Eq.~(\ref{PTdistEq}) which equals $\ln(2^n)-1 + \gamma$ with $\gamma$ representing the Euler's constant~\cite{boixo2018characterizing}.
\label{entropy20to26}\vspace{-18pt}}
\end{center}
\end{figure}

We calculate the entropy of the quantum distribution for qubits in the range of 22 to 28, for variable depths of a random quantum circuit, see Fig.~\ref{entropy20to26}.  We have indeed observed that as circuit depth increases, there is a general trend of increase in entropy, which reaches the maximal entropy of a Porter-Thomas distribution at the depth of quantum supremacy experiment implemented in \cite{supremacy2019quantum}. At circuit depth lower than 8, intriguingly, the entropy does not increase monotonically. This occurrence is preserved independent of qubit number.

\subsubsection*{Bit-wise Correlations}

\begin{figure}[ht]\vspace{-10pt}
\begin{center}
\includegraphics[width=0.6\linewidth]{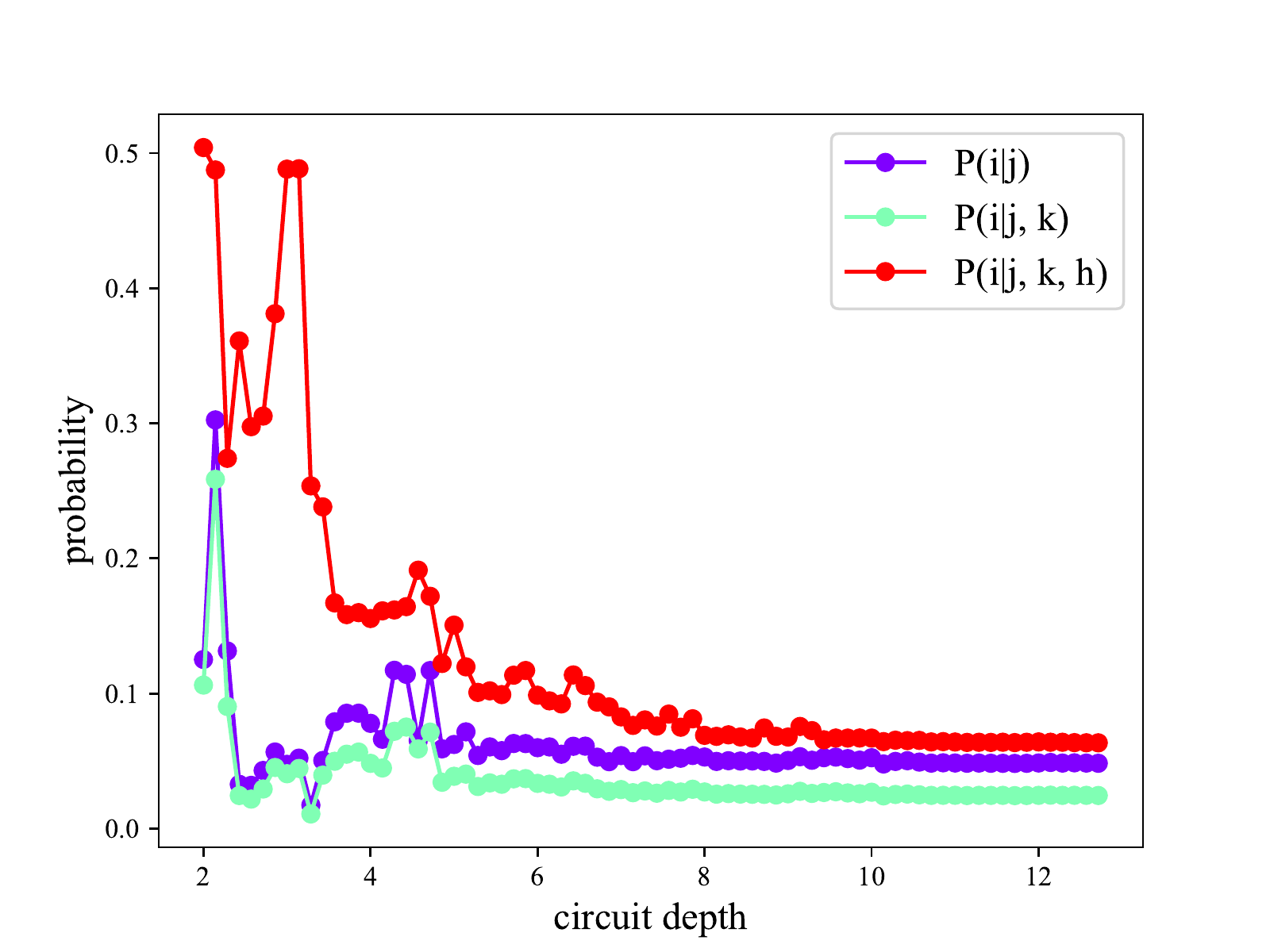}
\caption{The maximal conditional probability of one bit's value conditioned on one, two, and three qubits as a function of random quantum circuit depth.\vspace{-20pt} \label{maximalconditionalProbfig}
}
\end{center}
\end{figure}

\begin{figure*}[ht]
\begin{center}
\includegraphics[width=0.7\linewidth]{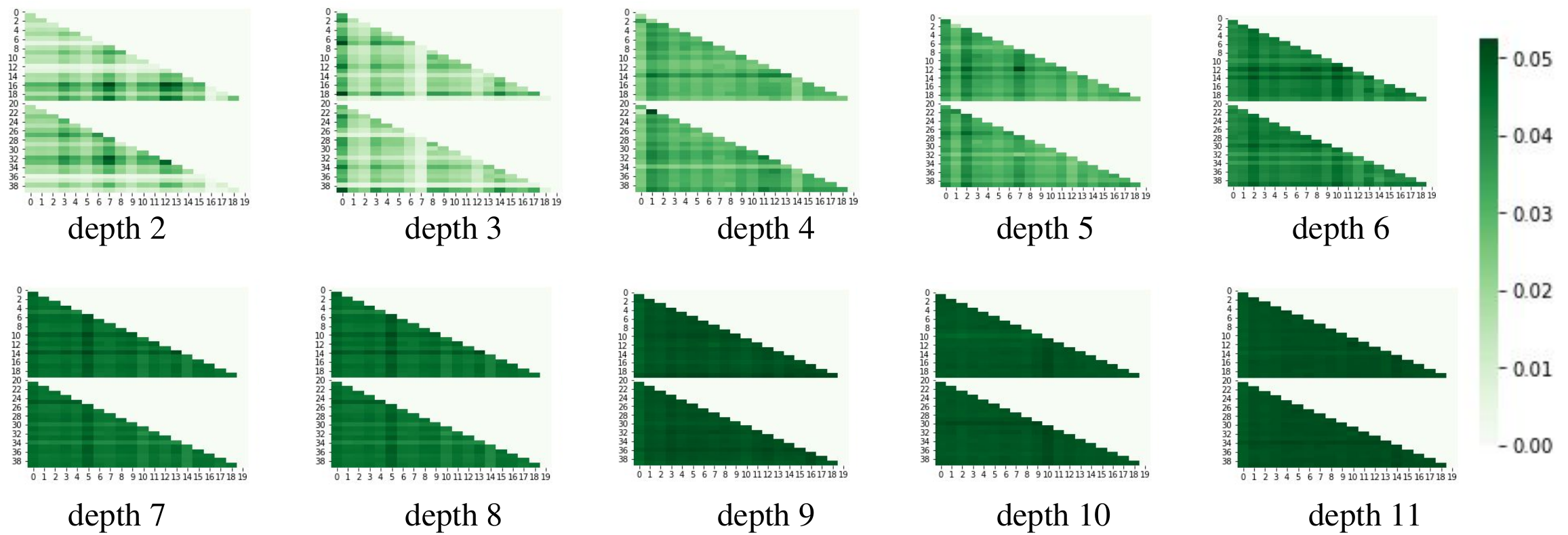}
\caption{The conditional probability of one bit's value conditioned on the second bit as a function of the position of the two bits in a bitstring, for variable depth of random quantum circuit from \cite{supremacy2019quantum} on 20 qubits.  As depth increases beyond 9, conditional probabilities approach to that of a i.i.d Porter-Thomas distribution where fine structures in bit-value correlations vanish.
\label{pairwiseConditionProb20QUBIT}}
\end{center}
\end{figure*}

The interesting structures unveiled by the entropy of the quantum distribution as a function of quantum circuit depth arises from oscillating correlations as a function of circuit depth. We probe the bit-wise correlations in greater detail by examining the conditional probability of the output as circuit depth varies. More specifically, we calculate the one-, two- and three-bit conditional distributions of the quantum distribution represented by:  $P(i=1|j=0)$ and $P(i=1|j=1)$ for pair-wise conditional probability of the $i$th bit being in state $1$ conditioned on the $j$th bit being 0 or 1, $P(i=1|j=0, k=0), P(i=1|j=1, k=0), P(i=1|j=0, k=1), P(i=1|j=1, k=1)$ for conditioning on two other bits; and simillarly for $\{P(i=1|j , k , h )\}$. The maximal conditional probability varies drastically as circuit depth increases from 1. Such drastic oscillation die down at depth around five, and quickly converges to that of a Porter-Thomas distribution at circuit depths above 8, see Fig.~\ref{maximalconditionalProbfig}. Such a transition coincides with the disappearing of fine structures in the entropy-circuit depth dependence. As is illustrated in the two-dimensional heat map of the pairwise conditional probability in Fig.~\ref{pairwiseConditionProb20QUBIT}, non-trivial spatial correlations arise at the low depth regime, and disappear as circuit depth increases beyond 8.

\section{Exact Representation of Random Quantum Circuit Probability Distribution}\label{app:dbm}
In this section,  we provide a method to use DBM for representing the probability distribution given by the random quantum circuit. 
\subsubsection{ Deep Boltzman Machine}\label{DBMSec}
The first method derives from deep Boltzman machines~(DBM). The basic idea is to represent complex-ising models with complex neural network of a given architecture ansazts: physical layers which consists of a layer of neurons each representing the physical degree of freedom $\{z_i\} \in [0,1]$, a hidden layer specified by neurons $\{h_j\}$, and a deep layer represented by neurons  $\{d_j\}$. We start from this minimalistic construction of a three-layer complex neural network, but later on we can increase the number of hidden layer and deep layer as we implement more and more complex quantum circuits. 

DBM seeks to represent the computational basis  $\ket{z}=\ket{z_1, \ldots, z_n }$ amplitudes for a generic quantum state $\ket{\psi}$ through the hidden and deep units of the neural network as:
\begin{align}
    &\bra{z} \psi \rangle = \Psi_{W}(z) =\sum_{h_i\in[1,-1], d_i\in[1,-1]  }e^{\left( \sum_i \omega_i z_i + \sum_{i,j}z_i W_{ij}h_j + \sum_j b_j h_j + \sum_{j,k}h_jd_k W_{j,k}^\prime + \sum_kb_k^\prime d_k \right)}
\end{align} 
where we use $W_{ij}$ to represent the weight matrices between the physical variables and first layer hidden variables, and $b_j$ as the baises of the same layer; we use $W^\prime_{j,k}$ to represent the weight matrices between the first hiden layer and the deep layer neurons, and $b_k$ the corresponding bias.

To find the DBM that corresponds to the random quantum circuit sampling amplitude for each bit string at the end of measurement, we can start from an identity empy circuit, and add DBM parameters iteratively. Let the $l$th cycle of quantum circuit be $U_1(l)U_2(l)$, assume we have already found the DBM parametrized by $W_{l-1}$ for the first $l-1$ cycles of quantum circuit, such that 
\begin{align}
      \bra{z}\left[\prod_{j=1}^{l-1} U_1(j)U_2(j)\right]H^n\ket{0^n}  = \Psi_{W_{l-1}}(z) ,
\end{align}
then to approximate the first $l$ depth of quantum circuit outcome probability distribution, we change $W_{l-1}$ to $W_l$ such that 
\begin{align}
      \bra{z}  U_1\ket{\Psi_{W_{l-1}}}  &= \Psi_{W_{l}}(z)=\sum_{h_{[lm]}}e^{z_l W_{l, [lm]}h_{[lm]} + z_m W_{m, [lm]} h_{[lm]} }\Psi_{W_{l-1}}(z),
\end{align} where the changes in the   neural network hidden layers $h_{[lm]}$~(or deep layers that intern affect the hidden layers) show up as a multiplicative factor and matches the muliplicative unitary transformation of the gate $U_l$.
We now discuss each DBM modification for constructing different $U_l$ that can be added sequentially to represent a quantum circuit specified by sequential unitary transformations.
\subsection{Initialization}

To initialize the DBM, we like to realize the outcome of measuring quantum state after an identity operation. This can be straightforwardly realized by adding one hidden variable for each physical variable~(which is the given qubit value):
\begin{align}
W_{l j}=\delta_{l,j}
\end{align}
\subsection{Single Qubit Gate Set}

A single-qubit gate from gate set $\mathbb{S}_1$ defined in Eq.~(\ref{singleqGateset1})-(\ref{singleqGateset3}) can be  represented in a general form by matrix exponentiation as
\begin{align}
    U_1(\theta,\phi) &=e^{-i\frac{\theta}{2}(\cos\phi \sigma^x  + \sin\phi \sigma^y )}\\
    \sigma^x& = \left(\begin{matrix}
    0 & 1\\
    1 & 0
    \end{matrix} \right)\\
    \sigma^y &= \left(\begin{matrix}
    0 & -i\\
    i & 0
    \end{matrix} \right)
\end{align} where $\sigma^x$ and $\sigma^y$ are two of the four independent generators for $SU(2)$ group, namely the Pauli X and Pauli Y operators. The gate set used for our random quantum circuit can thus be re-expressed by this gate as: $\sqrt{X} =U_1(\frac{\pi}{2}, 0), \sqrt{Y} =U_1(\frac{\pi}{2}, \frac{\pi}{2}), \sqrt{X+Y}  =U_1(\frac{\pi}{2}, \frac{\pi}{4})$.

The single-qubit gate  parameterized above
implements the following transformation on the measured amplitude in computational basis states:
\begin{align}\label{singlequbitNewDBMtarget}
    \bra{z} U_1(\theta,\phi, j)\ket{\Psi_{W }} =\cos(\theta)\Psi_W(z) + \sin(\theta)\Psi_W(z_j \to -e^{i\phi}z_j)=\Psi_{W_{new}}(z). 
\end{align}
According to this update rule, we specifically parameterize the new DBM $W_{new}$ as modifications on existing weights between physical layer and hidden layer $\Delta W{i,j}$, together with new deep variables $\{d_{[l]}\}$ each of which connects to a group of existing hidden variables $\{h_j\}$, and new hidden layer variables $h_{[l]}$. Let us also represents the previous DBM, namely $W$, by sump over  its hidden variables and deep layer variables $h$ and $d$ as
\begin{align}
    \Psi_W(z)= \sum_{h,d} P_1(z,h)P_2(h,d)P_3(d,z),
\end{align}where such factorized product form arises from the DBM ansatz: there is no intralayer connections, only interlayer connections between neurons. 
Then we have the following relation between the wave function amplitude of the new DBM with that of the old DBM:
\begin{align}
    \Psi_{W_{new}}(z)&=\sum_{h,d}\sum_{d_{[l]}}\sum_{h_{[m]}} P_1(z,h)P_2(h,d)P_3(d,z)e^{\sum_l \left[z_l \sum_j\Delta W_{l j}h_j + d_{[l]}\sum_j h_j W_{j[l]}^\prime +(z_l h_{[m]}W_{l,[m]} + h_{[m]}d_{[l]})  \right]}\\\label{singlequbitNewDBMfinal}
    & =\sum_{h,d}\sum_{d_{[l]}}  P_1(z,h)P_2(h,d)P_3(d,z)e^{\sum_l \left[z_l \sum_j\Delta W_{l j}h_j + d_{[l]}\sum_j h_j W_{j[l]}^\prime +(z_l  W_{l,[l
    ]}^{''} d_{[l]})  \right]}
\end{align} where the second equation comes from summing over the new hidden variables of this step which gives an effective weight matrices $W^{\prime\prime}$ directly between the deep layer and the physical layer. Equating the Eq.~(\ref{singlequbitNewDBMtarget}) with Eq.~(\ref{singlequbitNewDBMfinal}) we get the following solution for the modification for old DBM:
\begin{align}
 &   W_{j[l]}^\prime= - W_{l j} e^{i\phi}\\
 & \Delta W_{l j}  =  - W_{l j}\\
 &W_{j[l]}^\prime  = \frac{1}{2}\text{arcosh}\left(\frac{1}{\tan(\theta)}\right)
    \end{align}
    
In summary, to represent one single qubit gate from $\mathbb{S}_1$, we need to add for every pair of connection between physical and hidden layer, a new pair of    non-local connection between the physical layer and deep layer; and additionally a new hidden layer variable that connects to  both the newly added deep-layer variable and the physical variable the gate acts upon. Consequently, the number of hidden and deep layer variables grow linearly with the number of single qubit gate, but the number of weight matix entries grows exponentially with the number of single qubit gate.

\subsection{CZ gate}

In the second case, when the desired unitary evolution at the given time step is a two-qubit CZ gate on $i, j$ qubit pair represented by \begin{align}
    CZ(i,j) = e^{-i\pi(\sigma_i^z\sigma_j^z - \sigma_i^z - \sigma_j^z)}
\end{align}
We will focus on implementing the two-qubit evolution induced by $\sigma_i^z\sigma_j^z$ since the single qubit rotation corresponds to simple change in the weight factor of each physical layer and is trivially realizable. More generally, assume we like to find a new DBM which is equivalent to the original wavefunction represented by the old DBM mulitplied by a unitary transformation $U_{zz}(\psi, i, j)= e^{-i \psi \sigma_i^z \sigma_j^z}$. As is shown in ref.~\cite{carleo2018constructing}, this can be simply realized by adding one new hidden variable $h_{[i,j]}$ to the hidden layer which connects to physical variable $z_i$ and $z_j$ each with weights $W_{i,[ij]}$ and $W_{j,[ij]}$, such that summation over this new hidden variable induces desired correlation between the physical spins:
\begin{align}
    \Psi_{W_{new}}(z)&= \sum_{h_{[ij]}}e^{z_i W_{i,[ij]}h_{[ij]}+ z_j W_{j,[ij]}h_{[ij]}}\Psi_{W }(z).
\end{align}

If we choose the weight matrices for this new hidden variable as: 
\begin{align}
    W_{i,[ij]} = \frac{1}{2}\text{arcosh}(e^{ i 2 |\psi|})\\
    W_{j,[ij]} = \text{sgn}[\psi] \times W_{i,[ij]}
\end{align} then we obtained the desired transformation between the physical layers: $\Psi_{W_{new}}(z)  =  e^{-i\phi z_i  z_j}\Psi_{W }(z)$. In summary, every CZ gate introduce one more hidden layer, and two weight matrix connections to two physical variables that CZ is applied to.
 
\subsection{Two-Qubit fSim Gate}
Another two-qubit gate important for the random circuit is the fSim gate, which can be generically represented as a rotation induced by the anisotropic Heisenberg model. More specifically, let use represent the more general fSim gate on the $l$th and $m$th qubits by:
\begin{align}
    U_{\text{fSim}}(\theta, \phi, l,j)=e^{-i \theta(\sigma_l^x\sigma_m^x + \sigma_l^y\sigma_m^y)-i \phi\sigma_l^z\sigma_m^z }.
\end{align}
Without rederiving the detail, we present the DBM construction that realizes this update, described by the following steps:
\begin{enumerate}
    \item add a variable in the deep layer $d[lm]$ which is connected to all the hidden variables $h_j$ that connect to the physical spin $z_l$ and $z_m$ with a weight:
    \begin{align}
        W_{j[lm]}^\prime = W_{lj}- W_{mj}
    \end{align}
    \item delete all existing connection between physical spin $z_l$   and hidden variables $h_j$, and attach the hidden variables that connect only to spin $z_m$ with the weight that's identical to connection to $z_l$.
    
    \item Create a hidden variable $h_{[lm1]}$ that connects to the deep variable added in step 1 and the physical spin $z_l$ with weights:
    \begin{align}
        W_{l[lm1]}=W^\prime_{[lm1][lm]} = \frac{1}{2}\text{arcosh}[1/\tan(2\theta)]
    \end{align}
    
    \item Create a hidden variable $h_{[lm2]}$ that connects to both $z_l$ and $z_m$ with weight
    \begin{align}
        W_{l[lm2]} = -W_{m[lm2]} = \frac{1}{2}\text{arcosh}
\left(\cos(2 \theta)e^{i2\phi} \right)   \end{align}

\item Create a hidden variable that connects to both $z_l$ and $z_m$, and the deep variable $d_{[lm]}$ with weights:

\begin{align}
    W_{l[lm3]}=W_{m[lm3]}=-W{[lm3][lm]} =i\frac{\pi}{6 }
\end{align}

\end{enumerate}

\begin{figure}[ht]
\begin{center}
\includegraphics[width=0.5\linewidth]{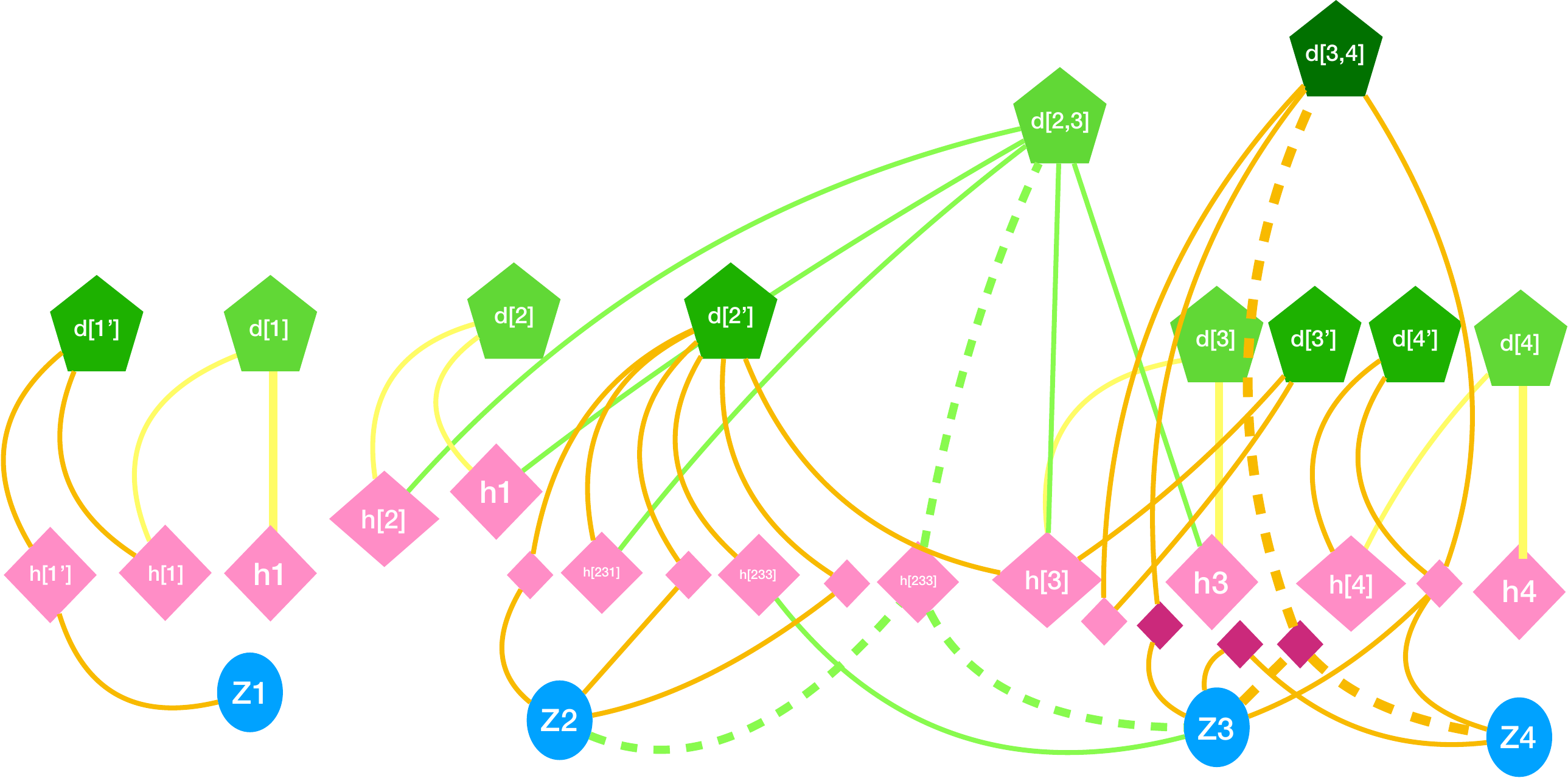}
\caption{A deep Boltzman Machine representation of depth 4 random circuit.
\label{DBMrepdepth4Circuit}}
\end{center}
\end{figure}

Following this principle, we provides an example for the depth 4 random circuit's DBM representation. The original circuit takes is represented by Fig.~1 in the main text. Its DBM is represented by Fig.~\ref{DBMrepdepth4Circuit}.
In summary, every fSim gate introduce one deep layer variable, three hidden variable, and the number of weight matrix connections proportional to existing hidden layer size. This means the total number of weight connections introduced by fSim gate scales exponentially in the number of gate, and therefore also in the depth of the circuit.

 Utilizing the explicit construction of DBM from random quantum circuit description, at initialization, we have $2n$ number of binary variables in total in DBM. By adding every layer of single qubit gate, the number of hidden variable doubles, and the number of deep variable increase by $n$. And assume that on average each qubit is touched at least by one fSim gate at each depth, then after a layer of two-qubit gates, the number of deep variable increases by $n$ and hidden variables increases by $3n$. As a results, the  total number of variables in both hidden and deep increases with circuit depth $d$ as $(6n)^d$. Based on the known sampling algorithm for DBM~\cite{carleo2018constructing}, the amount of sub-sampling steps over the hidden and deep variables scales as $2^N_{DBM}$, where $N_{DBM}\sim O((6n)^d)$ represents the total number of deep and hidden variables in DBM.
 
\section{GAN Experiment Specification}
We utilize four layers fully connected neural network for both generator and discriminator. The hyper parameters in learning rate for generator and discriminator, regularizers for generator and discriminator, as well as label smoothing for the last layer ResNet are optimized with automated hyper-parameter grid search. Implementation, see open source github link~\cite{open_source_code}.


\section{SeqGAN Details}
We use the reference open-source implementation of SeqGAN with the LSTM generator hidden size set to 256, 512 and 1024. We use a batch size of 64, learning rate of 1e-4 and embedding dimension 32. We use the default CNN discriminator with filter sizes of 2, 3, 4, 5, 6, 7, 8, 9, 10, 15 and 20 with a total of 1620 filters and with any filters longer than the sequence length removed. The output of the CNN is fed into at 256 dimension hidden layer which is then used to predict whether the input sequence is from the training distribution or the generated distribution. We do not perform maximum likelihood pretraining since this is a binary prediction task and so we do not expect pretraining to be necessary.

\section{LSTM Hyperparameter tuning}
We tuned the LSTM learning rate over several orders of magnitude including 0.1, 0.01, 0.001, 0.0001 and 1e-5 and used 0.001 which consistently performed the best over multiple numbers of qubits. We train using the Adamax optimizer which we found to consistently perform better than Adam and used a fixed batch size of 64. We apply no regularization or dropout. The output of the LSTM is fed into a logistic layer. We provide the source code for our implementation of LSTM in files included in the submission folder: $ \textit{data\textunderscore loader}, \textit{\textunderscore init \textunderscore .py}, \textit{run \textunderscore lm.py}, \textit{run.sh}$.

\section{Generate Quantum Samples from Experiments with Laboratory Quantum Computers}

Quantum samples consist of a set of $n$-bit strings from a given random quantum circuit on $n$ qubits can be obtained by applying the random quantum circuit to all zero states $\ket{0}^{\otimes n}$, and then perform the quantum measurements. The outcome of each such quantum measurementis an $n$-bit string drawn from a   probability distribution determined both by the perfect quantum circuit unitary and  noise~(approximated by a uniform distribution) caused by imperfect realistic execution of quantum circuit.  The number of bit strings obtained   therefore equals the number of such repetitions. An example data set on the bitstrings samples obtained on 12 qubit experiments is included in the submission data folder, named $\textit{experimental \textunderscore samples \textunderscore q12c0d14.txt}$.

\section{Generate Quantum Samples from Theoretically Simulated Quantum Distributions}

For random circuit with qubit numbers smaller than 20, quantum library Cirq~\cite{cirq} simulator \textit{cirq.simulator} is sufficient to estimate the probability distribution for a random quantum circuit defined in \citet{supremacy2019quantum}. We included a python file named \textit{circuit.py} in the folder, which specifies the 12 qubit circuit we have studied, and code for simulating the outcome probability. This code utilizes the \textit{cirq.simulator}, which takes a circuit defined in \textit{cirq.circuit()} form as input~(specified in the beginning of \textit{circuit.py} file), and output the amplitudes of the wavefunciton for each one of $2^n$ computational states, whose absolute values squared is the probability of the corresponding bit string.

For random circuit with qubit number larger than 20, we adopt the  qsim simulator in \cite{supremacy2019quantum}, which is a
Schr\"{o}dinger full state vector simulator. It computes all $2^n$ amplitudes, where $n$ is the number of qubits. Essentially, the simulator performs matrix-vector multiplications repeatedly. One matrix-vector multiplication corresponds to applying one gate. For a 2-qubit gate acting on qubits $q_1$ and $q_2$, it can be depicted
schematically by the following pseudocode in Algorithm~\ref{alg1}, see open-sourced code at~\citep{qsim}.

\begin{algorithm} 
\caption{Classical simulation of quantum circuit output wavefunction}\label{alg1}
  \begin{algorithmic}[1]
    \Procedure{QSim}{$q_1, q_2, U$}
      \For{\texttt{(int $i = 0; i < 2^n; i += 2 \times 2^{q_2}$)}}\Comment{iterate over all values of qubits $q > q_2$}
      \For{\texttt{(int $j = 0; j < 2^{q_2}; j += 2 \times 2^{q_1}$)}}\Comment{iterate values for $q_1 < q < q_2$}
      \For{\texttt{(int $k = 0; k < 2^{q_1}; k += 1$)}}\Comment{iterate values for $q < q_1$}
        \For{\texttt{  all $q\notin [q_1,q_2]$}}\Comment{iterate values for $q < q_1$}
        \State \texttt{int $l = i + j + k;$}
\State \texttt{float $v0[4]$;} \Comment{gate input}
\State \texttt{float $v1[4]$;} \Comment{  gate output}
\State \texttt{$v0[0] = v[l]$;}\Comment{copy input}
\State \texttt{$v0[1] = v[l + 2^{q_1}]$;}\Comment{copy input}
\State \texttt{$v0[2] = v[l +  2^{q_2}] $;}\Comment{copy input}
\State \texttt{$v0[3] = v[l + 2^{q_1} + 2^{q_2}]$;}\Comment{copy input}
\For{\texttt{ ($r = 0; r < 4; r += 1$)}}\Comment{apply gate}
\State \texttt{$v1[r] = 0$;}
\For{\texttt{ ($s = 0; s < 4; s += 1$)}} 
\State \texttt{$v1[r] += U[r][s] * v0[s]$;}
 \EndFor
 \EndFor
\State \texttt{$ v[l] = v1[0]$;} \Comment{copy output}
\State \texttt{$v[l + 2^q_1] = v1[1]$;}\Comment{copy output}
\State \texttt{$v[l + 2^q_2] = v1[2]$;}\Comment{copy output}
\State \texttt{$v[l + 2^q_1 + 2^q_2] = v1[3]$;}\Comment{copy output}
      \EndFor
      \EndFor
      \EndFor
      \EndFor
      \State \texttt{\textbf{return} $v$ }
    \EndProcedure
  \end{algorithmic}
\end{algorithm}

\section{Random Porter-Thomas Learning}
\begin{figure}[h!]
\begin{center}
\includegraphics[width=0.6
\linewidth]{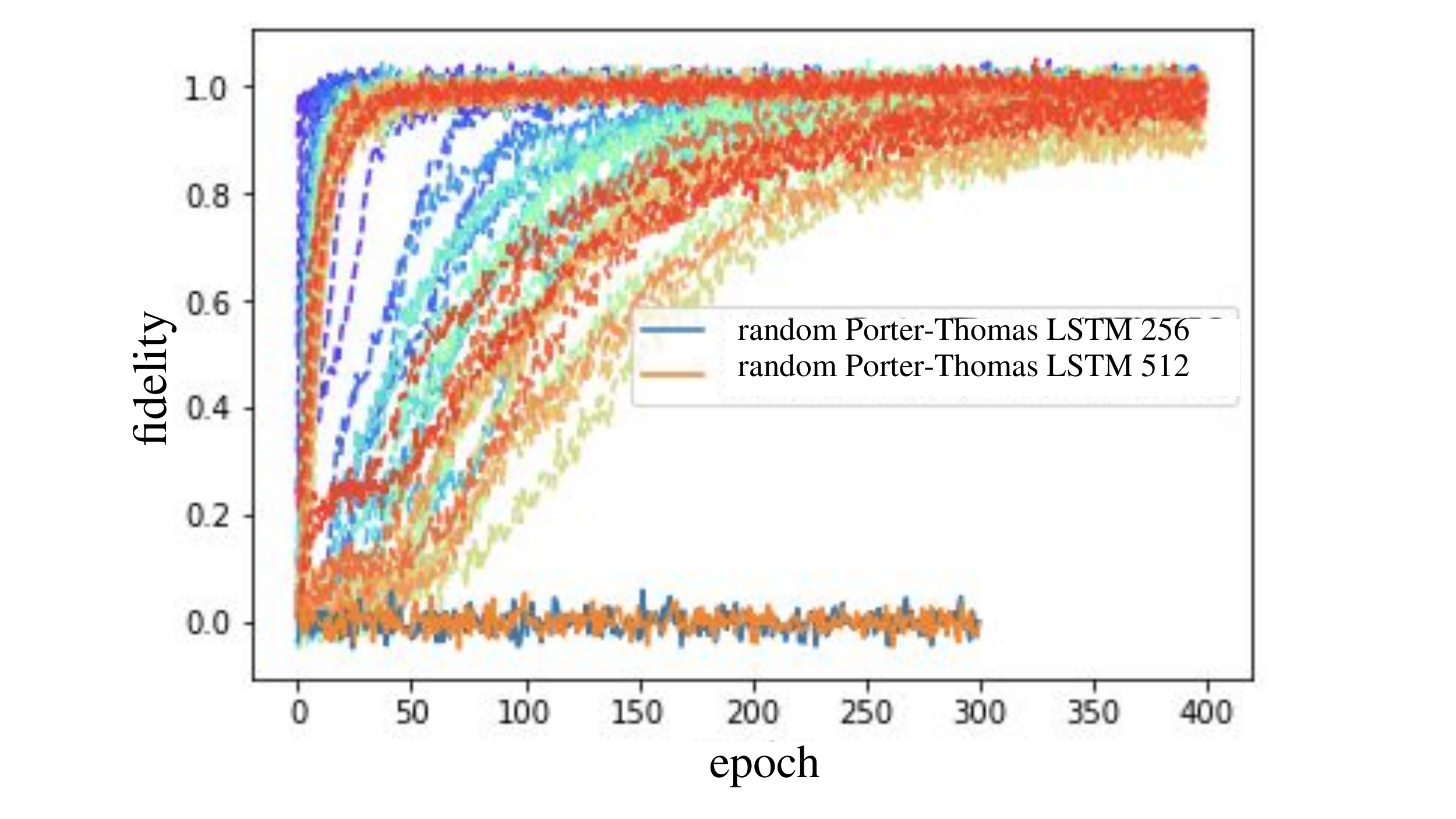}
\caption{The learned fidelity as a function of training epochs, dashed lines are for $m$-bit subset parity re-ordered Porter-Thomas learning with LSTM of size 256, with $m\in[2,4,\ldots, 28]$. Solid orange and blue line represents the learned fidelity for Porter-Thomas over randomly ordered $28$-bit strings. The fidelity for $m$-bit subset parity re-ordered Porter-Thomas  learning eventually approach unity, while in constrast, the learned fidelity of random Porter-Thomas remain close to zero despite given similar amount of training epochs.
\label{randomPT28}}
\end{center}
\end{figure}

The performance of LSTM learning randomly ordered Porter-Thomas over 28-bit strings is shown in Fig.~\ref{randomPT28}: it's close to zero during the first 300 epochs of training, while in comparison, the same training on $m$-bit subset parity re-ordered Porter-Thomas achieves $>0.8$ fidelity for all $m\leq 28$.

 \section{$\chi^2$ Test}

Given $N$  random $n$-bit  samples from a given quantum distribution, there are $2^n$ unique classes labeled by the corresponding $n$-bit strings. A null hypothesis assigns the probability $p_i$ for $i$th class with $i\in\{1, 2, \ldots, 2^n\}$. So we have the expected numbers $m_i = N p_i$ many $i$th bitstrings, with $\sum_i m_i=N$. We also have samples from a given observation $\{x_i\}$, for example generated by our trained generative model, satisfying $x_i\in \{1, 2, \ldots, 2^n\}$ and $\sum_i x_i=N$. $\chi^2$ test calculate the following value:
\begin{align}
  \chi^2=  \sum_{i=1}^{2^n}\frac{x_i-m_i}{m_i},
\end{align}
which itself obeys normal distribution as $N$ goes to infinity. By calculating $ \chi^2$, we can obtain the p-value of the null hypothesis. A small p-value means the probability that the given samples $\{x_i\}$ are drawn from the given null-hypothesis distribution is low. When $\chi^2=0$, we have p-value reaches the maximum of 1. 

\begin{figure}[h!]
\begin{center}
\includegraphics[width=0.5\linewidth]{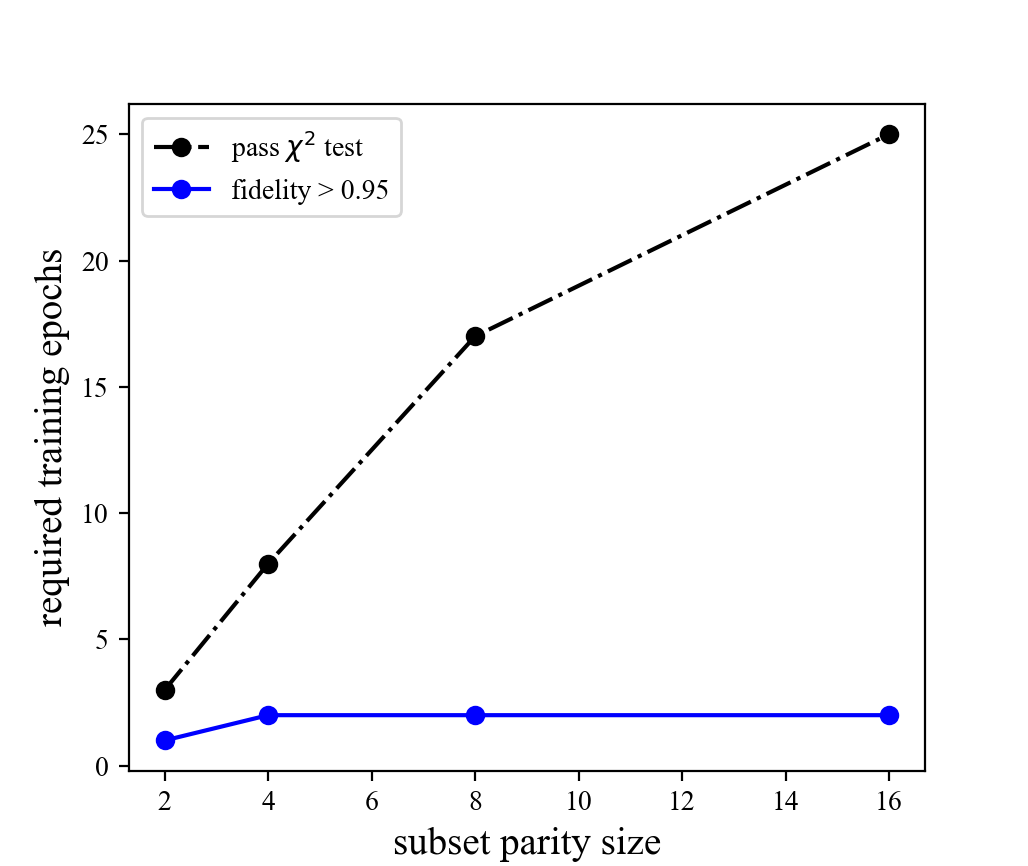}
\caption{Required training epoch to reach a given
\label{chi2fidelitycompare}}
\end{center}
\end{figure}

\begin{figure}[h!]
\begin{center}
\includegraphics[width=0.7\linewidth]{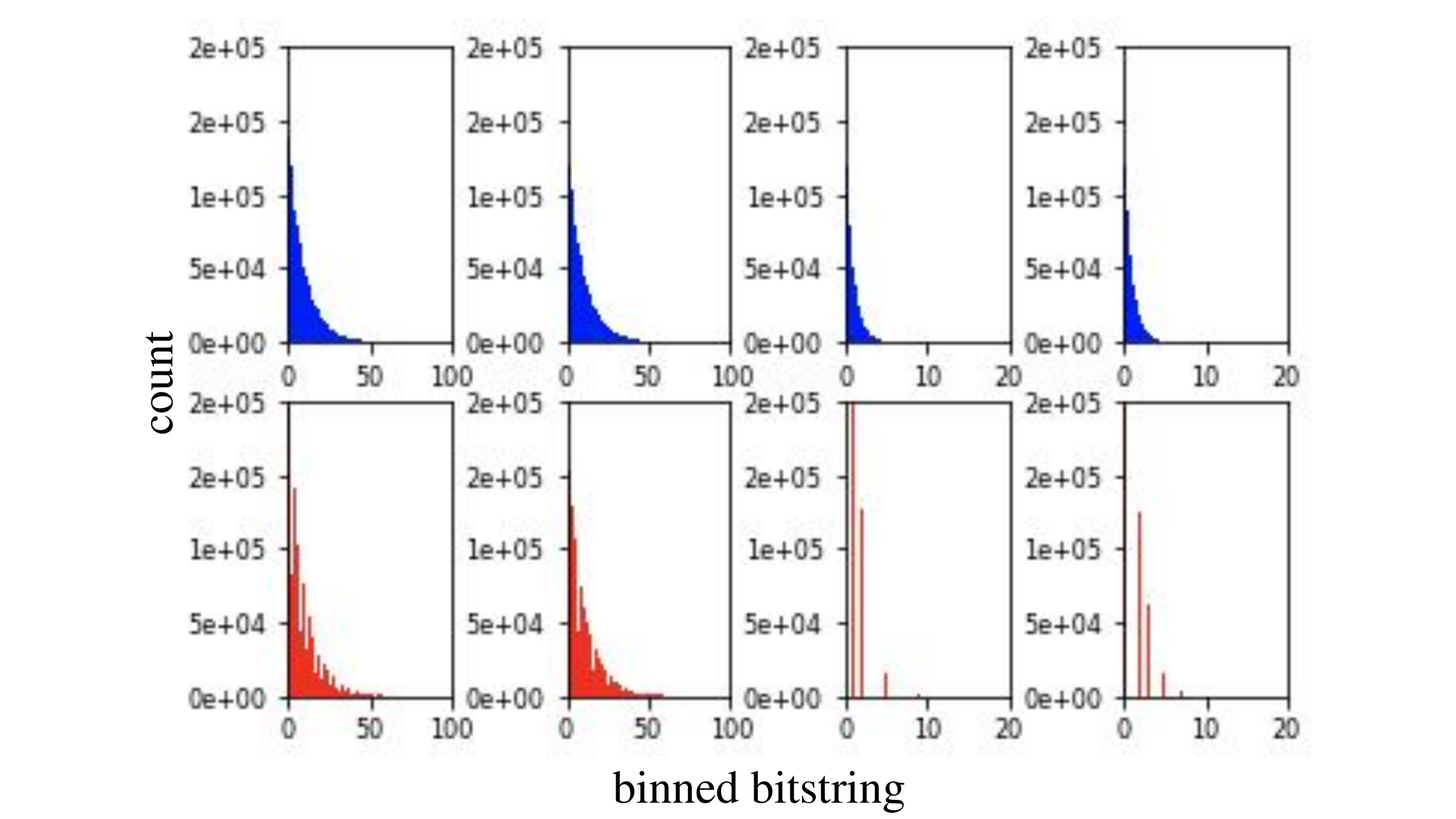}
\caption{Histogram of one million samples over all possible bitstrings drawn from the true quantum distribution~(blue plots in the upper row), vs drawn from the learned distribution~(red plots in the second row). Different columns corresponds to subset parity of different number of bits. From the leftmost to rightmost column, it corresponds to subset parity of bits ranging from 2, 4, 8, to 16.
\label{chi2histogram}}
\end{center}
\end{figure}

Compared to fidelity definition if Eq.~(6), $\chi^2$ is  a more robust measure in that it is sensitive to the shape of the empirical distribution. In comparison, one can fool the fidelity measure by replacing all samples with the same bitstring which has the maximum probability in the null hypothesis. For example, the fidelity measure for learning subset parity re-ordered Porter-Thomas for 20 qubits reaches unity in merely 2 epochs for $n$-bit subset parity with $n<=20$, but to pass the  $\chi^2$ test, the required  number of training epochs increases quickly with $n$, see Fig.~\ref{chi2fidelitycompare}.  The difference between fidelity and $\chi^2$ measure is even more salient in the shape of empirical distribution: Fig.~\ref{chi2histogram} shows the histogram of the empirical distribution of LSTM model in the final epoch, for $n$-bit subset parity reordered Porter-Thomas distribution, with $n=2, 4, 8$ and $16$. While the fidelity of all four cases are close to 1 in the last epoch, the p-value from $\chi^2$ gives $0.05$ for $n=16$, which discrepancy from theoretical distribution is visible in  the right most column of Fig.~\ref{chi2histogram}.

\bibliography{main}
\end{document}